\documentclass[]{aa} 
\usepackage[colorlinks=true, linkcolor=blue, citecolor=blue, urlcolor=blue]{hyperref}
\usepackage{amsmath}
\usepackage{graphicx}
\usepackage{amssymb}
\usepackage{mathrsfs}
\usepackage{caption}
\usepackage{float}
\usepackage{afterpage}
\usepackage{amstext}
\usepackage{multirow}

\usepackage{txfonts}
\usepackage{natbib}

\defcitealias{Tautvaisiene2020}{Paper~I}
\defcitealias{Tautvaisiene2021}{Paper~II}
\defcitealias{Tautvaisiene2022}{Paper~III}
\begin{document}

   \title{Calibration of the [C/N] and [Y/Mg] chemical clocks with asteroseismic ages from the TESS space mission
}
   \author{
E. Pak\v{s}tien\.{e}\inst{1},
G. Tautvai\v{s}ien\.{e}\inst{1}, 
V. Bagdonas\inst{1},
H. Kjeldsen\inst{2},
M. L. Winther\inst{2},
A. Drazdauskas\inst{1},\\
C. Viscasillas V\'azquez\inst{1},
Y. Chorniy\inst{1},
\v{S}. Mikolaitis\inst{1},
R. Minkevi\v{c}i\={u}t\.{e}\inst{1},  
E. Stonkut\.{e}\inst{1}
}

   \institute{Institute of Theoretical Physics and Astronomy, Vilnius University, Saul\.{e}tekio av. 3, LT-10257 Vilnius, Lithuania\\
              \email{erika.pakstiene@tfai.vu.lt } 
              \and
    Stellar Astrophysics Centre, Department of Physics and Astronomy, Aarhus University, Ny Munkegade 120, 8000 Aarhus, Denmark 
             }

   \date{Received 8 January 2026 / Accepted 12 February 2026}

\authorrunning{E. Pakštienė et al.}
\titlerunning {Calibration of the [C/N] and [Y/Mg] chemical clocks with asteroseismic ages}

 
  \abstract
   {Stellar ages are typically very difficult to estimate for field stars. New empirical methods, based on abundance ratios of chemical elements, are emerging and need to be calibrated. 
   }
   {Our main aim is to contribute to revealing relations between [C/N] and [Y/Mg] ratios and stellar ages by determining astroseismic ages and non-local thermodynamic equilibrium (NLTE) abundances, and accounting for stellar evolutionary stages and birth places in the Galaxy. 
}
   {We searched for solar pulsations in a sample of 1250 bright F, G, and K giants using data from the TESS space telescope and determined asteroseismic ages using the BASTA and PARAM codes. For the [Y/Mg] relations with age, we determined abundances accounting for deviations from the local thermodynamic equilibrium. For the [C/N] relations with age, we separated stars according to their evolutionary stages.  } 
   {We determined asteroseismic ages for 218 giants and derived [Y/Mg] and [C/N] relations with age for subsamples of stars in three regions of the Galactic thin disc and the thick disc.  }
  {The [Y/Mg]--age relation exhibits a clear radial dependence
across the Galactic disc, with a steeper trend in the outer disc,
progressively flatter relations towards the inner disc, and a very flat trend in the thick disc. NLTE abundances of Mg and especially of Y have to be used in order to obtain a more precise stellar age evaluation from [Y/Mg] ratios. When using [C/N] abundance ratios as stellar age indicators, evolutionary stages of stars have to be taken into account. 
  }


   \maketitle

\section{Introduction}

Age, a parameter describing how much time has passed since the birth of any object or since a specific event took place, is one of the most important parameters in astronomy. In order to paint the picture of the evolution of the Universe, it is crucial to constrain ages and timescales of basically anything that has ever existed. Only then can we properly investigate the nature of which we are part.

Unfortunately, precise ages of objects and events in the Universe are debatable in relation to their fundamentals. The best example is probably the age of the Universe, or, in other words, how much time has passed since the Big Bang. With the help of advancing technology, the \cite{Planck20} has managed to set the age of the Universe at 13.797 billion years. However, there are studies proposing a much older age (e.g. \citealt{Gupta2023,Llorente2024}). The Milky Way has its own time-related mysteries as well. For example, it is still unclear when its main structural elements –thin and thick discs– formed \citep{Kilic17, Xiang22}.

The best way to determine the ages of larger structures is to investigate its constituents, in this case the stars. However, many shortcomings do exist here as well. In stellar physics, the most common age determination methods have always been isochrone based, yet, even with most modern tools, these methods provide results with quite large uncertainties, sometimes spanning several billion years, making precise galactic investigations impossible \citep{Soderblom10}. 

Another alternative for determining stellar age is based on asteroseismology \citep[e.g.][]{Aerts21}. In this method, stellar oscillations of various stars are used to infer stellar ages. However, it has its own shortcomings -- stellar oscillations cannot be detected in all stars. Nevertheless, this method is much more precise than others, and since the start of the NASA Transiting Exoplanet Survey Satellite (TESS, \citealt{Ricker15}) space mission, covering almost all the sky, tens of millions of light curves have been collected; thus, ages determined from asteroseismology will continue to grow.

During the last decade, a new empirical method for the age determination has been extensively studied. This so-called chemical-clock method relies on astrospectroscopy, in particular, on stellar chemical abundances and their ratios, some of which have been noticed can, more or less, indicate stellar ages (e.g. \citealt[]{daSilva2012, Salaris15, Nissen15, Nissen16, Spina16, TucciMaia16, Adibekyan16, Lagarde17, Slumstrup17, Spina18, Casali19, DelgadoMena19,  Titarenko19, Casali20, Jofre20, Tautvaisiene2021, Spoo22, Vazquez22, Shejeelammal24, Vitali2024, Molero25, Roberts25, Tautvaisiene2025}). However, this empirical method is not trivial as abundances of various chemical elements found in stars are a function of many known and still unknown parameters. Abundance ratios are noted to be different in objects of different ages, metallicities, and galactic structural components \citep{Masseron2015, Casali20, Tautvaisiene2021, Vazquez22, Roberts25}. 

Presently, the best identified chemical-element ratios used in chemical-clock methods are of $s$- and $\alpha$-process origins, especially [Y/Mg] \citep{Nissen15, Casali20, Tautvaisiene2021, Vazquez22}. The reason for this is that the $s$ process, while responsible for a wide range of elements and being of rather different origins, mostly occurs on longer timescales, and its effect is most apparent in younger objects. In contrast, enrichment of the interstellar medium with $\alpha$ elements progressed on much shorter timescales. Therefore, the ratio of [$s$/$\alpha$] shows a prominent relationship with age. However, the calibration of this ratio as a chemical clock requires further efforts and large samples of different objects. 

Recent studies also demonstrated that ratios of mixing-sensitive chemical elements, carbon and nitrogen, can also be used in age inference; however, this is only for stars after the first dredge-up (1DUP) \citep{Salaris15, Martig16, Lagarde17, Casali19, Spoo22, Tautvaisiene2025}. It was also noticed that evolution stages of  stars after the 1DUP have to be taken into account as well (\citealt{Tautvaisiene2025}, and references therein) since C and N abundances change during the path of stars through its evolutionary track. The first changes occur during the 1DUP \citep{Iben65}, when in the low-mass stars the isotopes of $^{13}$C and $^{14}$N are brought to the surface, while $^{12}$C diffuses inwards, resulting in the reduction of $^{12}\text{C}/^{13}\text{C}$ and C/N ratios. In stars with masses below $\sim 2.2\,M_\odot$, these ratios decline further after the red giant branch (RGB) luminosity bump because of the extra-mixing processes start acting and last at least until the tip of the RGB (e.g. \citealt{Lagarde19}). Thus, the different behaviour of C and N abundances may serve as stellar age indicators in red giants.

The current situation is that both [Y/Mg] and [C/N] show potential to be useful age indicators; however, their age relations are not confined enough with respect to galactic locations, metallicities, and other parameters. The ability to determine ages from asteroseismology will allow us to define various chemical-clock relations with higher precision. Thus, asteroseismic ages and high-resolution spectral abundance investigations are needed.

In this paper, we present ages determined using asteroseismic data from the TESS space telescope. In our study, we investigated [Y/Mg] and [C/N] ratios as chemical clocks, taking into account stellar evolutionary phases, mean galactocentric distances, and thin- and thick-disc attributions.

\section{Stellar sample and method of analysis} 

In this study, we targeted 1250 F, G, and K spectral type stars with $V < 8$~mag in a field of about 45~deg radius centred on the TESS continuous viewing zone (CVZ) in the Northern Hemisphere. The stellar atmospheric parameters and abundances of chemical elements for this study were taken from \citet{Tautvaisiene2020}, \citet{Tautvaisiene2021}, and \citet{Tautvaisiene2022}, hereafter called \citetalias{Tautvaisiene2020}, \citetalias{Tautvaisiene2021}, and \citetalias{Tautvaisiene2022}. Only yttrium abundances for 110 stars had to be determined additionally in this work using the archival stellar spectra of the same \citetalias{Tautvaisiene2020} and \citetalias{Tautvaisiene2022}, and the ages were determined using data from the TESS space mission.  

The observations in \citetalias{Tautvaisiene2020} and \citetalias{Tautvaisiene2022} were carried out with the Vilnius University Echelle Spectrograph (VUES) \citep{2014SPIE.9147E..7FJ, Jurgenson16} mounted on the Moletai Astronomical Observatory $f/12$ 1.65~m Ritchey--Chretien telescope. The VUES is a multi-mode spectrograph that covers a wavelength range from 4000 to 8800~\AA. A spectral resolution mode with $R \sim 68\,000$ was used for observations of M spectral-type stars, and a mode with $R \sim 36\,000$ was used for other objects. The list contains no double-line spectroscopic binaries or stars rotating faster than 25~km\,s$^{-1}$. 

\subsection{Galactocentric distances and isochronal ages} 
\label{sec:distances and ages}

\begin{figure}
    \centering
    \includegraphics[width=0.47\columnwidth]{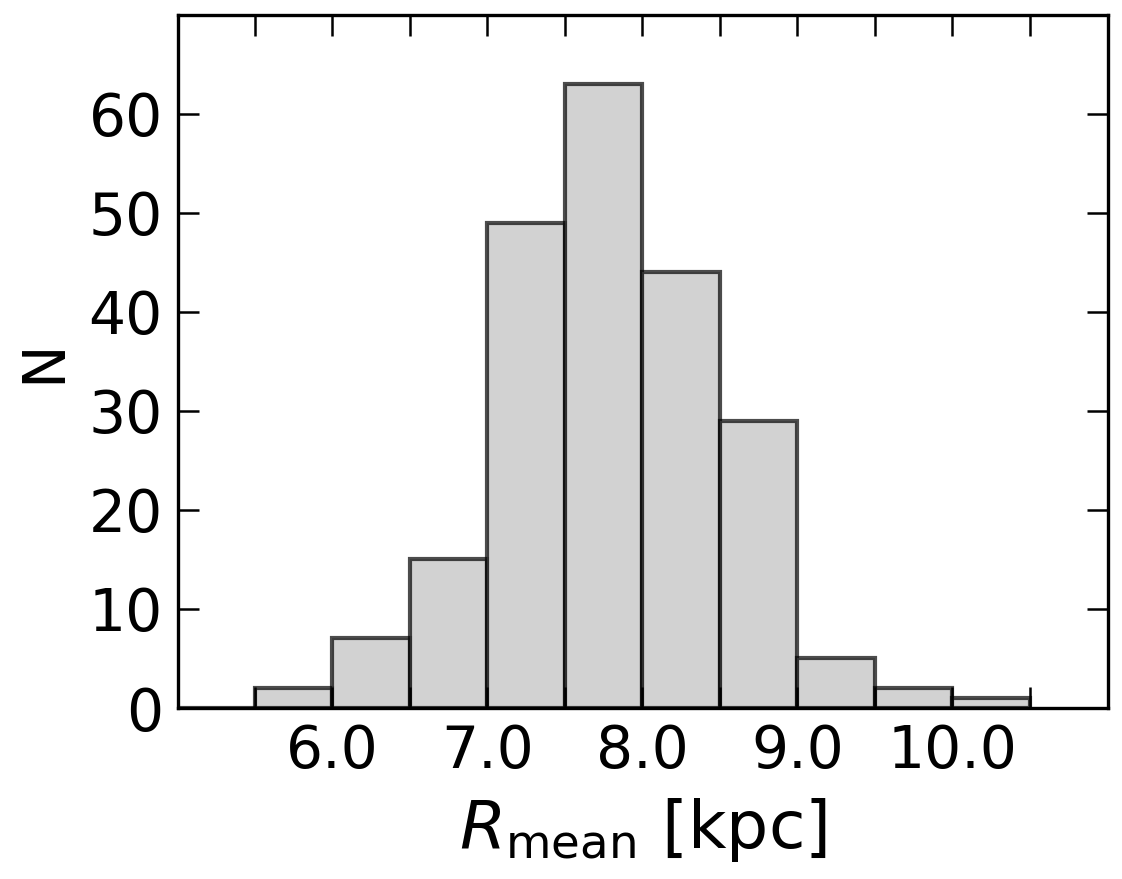}
    \includegraphics[width=0.47\columnwidth]{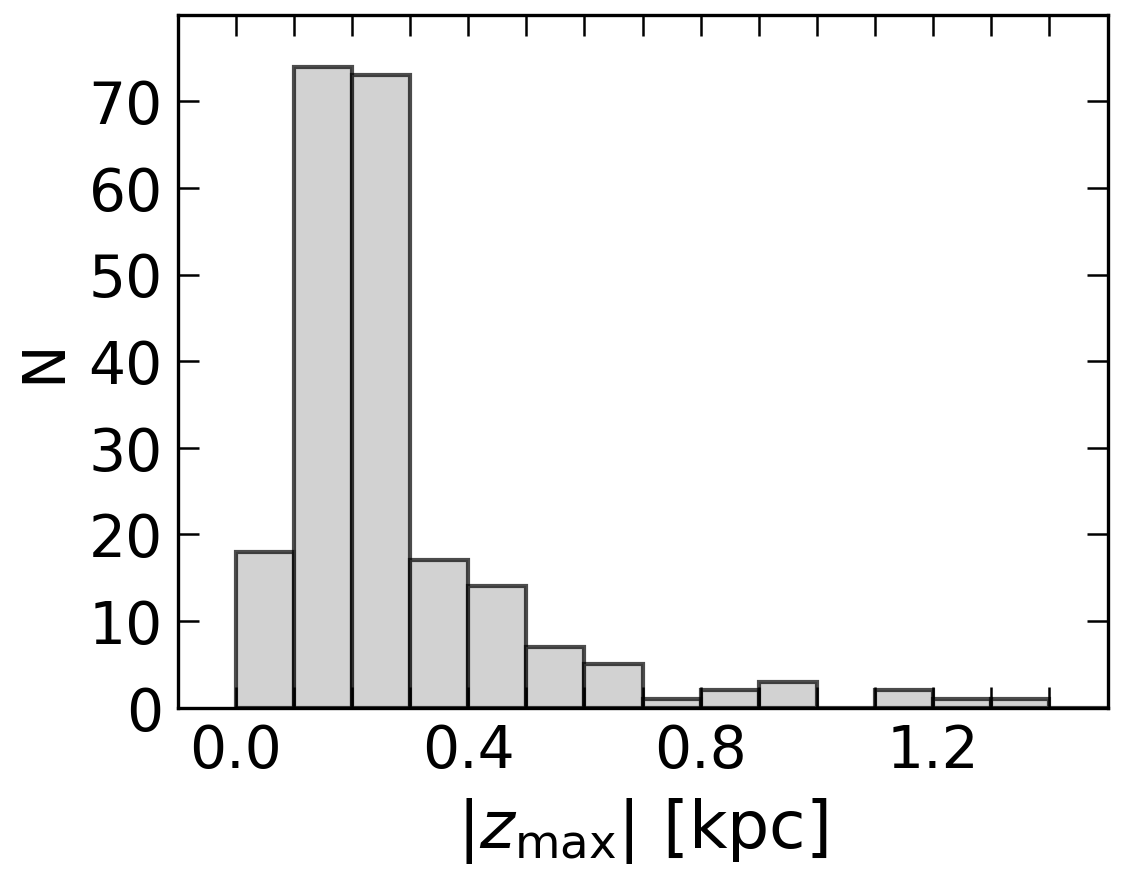}
    \caption{Distribution of stars according to mean galactocentric distance, $R_{\rm mean}$, and maximum height from the Galactic plane, |$z_{\rm max}$|.}
    \label{histo-kinem}
\end{figure}

In this study, we also used the mean galactocentric distances ($R_{\rm mean}$), the maximum distance from the Galactic plane (|$z_{\rm max}$|), and the isochronal ages taken from \citetalias{Tautvaisiene2020} and \citetalias{Tautvaisiene2022}. 
In these studies, the kinematic parameters were computed using the Python-based package $galpy$ \citep{Bovy15} by integrating orbits for 5~Gyr within the default MWPotential2014. The Solar parameters of $R_\odot = 8$~kpc and $V_{circ} = 220$~km\,s$^{-1}$ \citep{Bovy12}, with a distance from the Galactic plane of $z_\odot = 0.02$~kpc \citep{Joshi07} and the solar motion relative to the LSR adopted from \cite{Schonrish10}. While \citetalias{Tautvaisiene2020} used the input data from $Gaia$ DR2 \citep{Luri18, Katz2019, Gaia16, Gaia18}, \citetalias{Tautvaisiene2022} used the photogeometric distances from \cite{Bailer21} alongside the radial velocities of $Gaia$ DR3 \citep{Gaia16, Gaia18, Lindegren21, Seabroke21}. In both cases, uncertainties were estimated via 1000 Monte Carlo calculations,  taking into account uncertainties in the input parameters. 
Figure~\ref{histo-kinem} shows the distributions of the investigated stars according to their $R_{\rm mean}$ and |$z_{\rm max}$|.

\begin{figure}
    \centering
    \includegraphics[width=0.47\columnwidth]{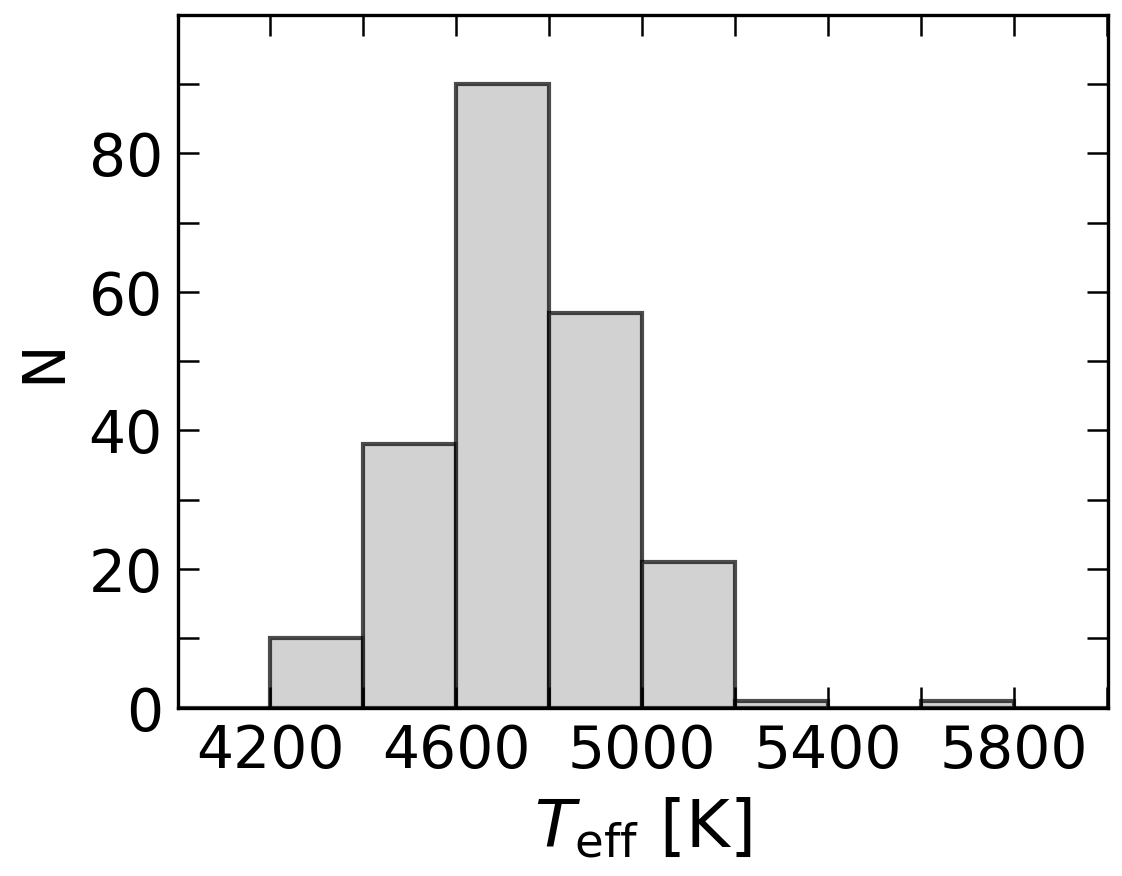}
    \includegraphics[width=0.47\columnwidth]{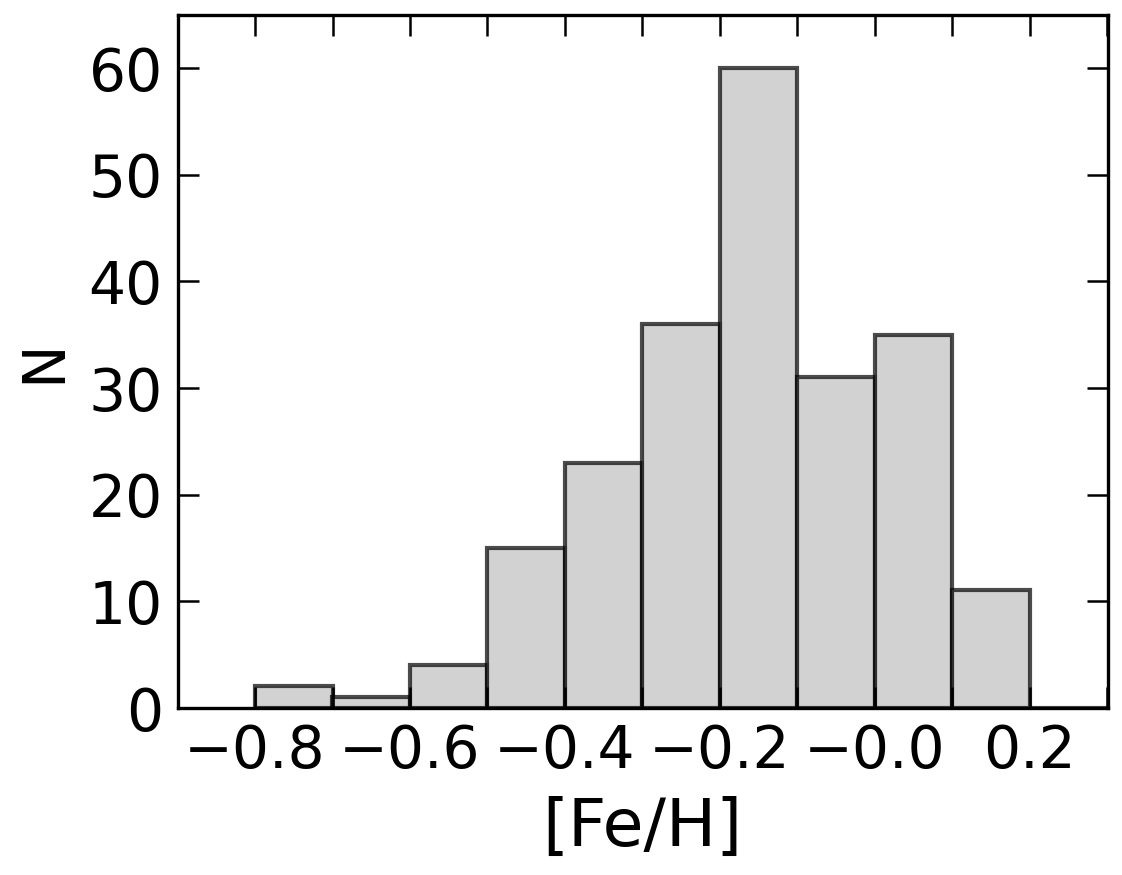}   
    \caption{Distribution of stars according to $T_{\rm eff}$ and [Fe/H].}
    \label{histo-atmos}
\end{figure}

The isochronal ages were used in this work for comparison only. They were also taken from \citetalias{Tautvaisiene2020} and \citetalias{Tautvaisiene2022}, in which they were  determined with a UniDAM code by \citet{2017A&A...604A.108M}.
This code combines spectroscopic data with infrared photometry from 2MASS \citep{Skrutskie06} and AllWISE \citep{Cutri14} and compares them with PARSEC \citep{Bressan2012} isochrones to derive probability density functions (PDFs) for ages. The maximum error allowed for age was 3~Gyr. The mean uncertainty of the age determination calculated from the uncertainties of individual stars was $\sim 2$~Gyr, the mean uncertainty of the $R_{\rm mean}$ determinations was $\sim 0.2$~kpc, and the mean uncertainty of the |$z_{\rm max}$| determinations was $\sim 0.05$~kpc.

\subsection {Stellar atmospheric parameters and abundances of chemical elements} 
\label{sec:chemistry}

The stellar atmospheric parameters (effective temperature $T_{\rm eff}$, surface gravity log\,$g$, metallicity [Fe/H], and microturbulent velocity $v_{\rm t}$) in \citetalias{Tautvaisiene2020} and \citetalias{Tautvaisiene2022} were determined from the equivalent widths of Fe\,{\sc i} and Fe\,{\sc ii} lines using standard spectroscopic techniques. The effective temperatures were derived by minimising the slope of the abundances determined from Fe\,{\sc i} lines with different excitation potentials. Surface gravities were found by requiring Fe\,{\sc i} and Fe\,{\sc ii} lines to give the same iron abundances. The microturbulent velocity values were attributed by requiring Fe\,{\sc i} lines to give the same iron abundances regardless of their equivalent widths. In total, 299 Fe\,{\sc i} and Fe\,{\sc ii} lines were used to calculate the stellar atmospheric parameters with the tenth version of the MOOG code \citep{1973PhDT.......180S} and the MARCS\footnote{http://marcs.astro.uu.se/} grid of constant flux local thermodynamic equilibrium (LTE) stellar atmosphere models \citep{Gustafsson08}. The calculated medians of atmospheric parameter determination errors from all the stars in our sample are the following: $\sigma T_{\rm eff}=\pm46$~K; and $\sigma {\rm log}\,g=\pm0.30$, $\sigma {\rm [Fe/H]}=\pm0.11$, and $\sigma v_{\rm t}=\pm0.27$~km\,s$^{-1}$. Figure~\ref{histo-atmos} shows the distribution of stars according to $T_{\rm eff}$ and [Fe/H].

Abundances of carbon, nitrogen, magnesium, and yttrium were also taken from \citetalias{Tautvaisiene2020}, \citetalias{Tautvaisiene2021}, and \citetalias{Tautvaisiene2022}; however, additional abundance determinations for yttrium were made for 110 stars using the same method of analysis.  
Abundances of the investigated chemical elements were determined using spectral synthesis. 
For modelling of synthetic spectra, we used a spectrum synthesis code Turbospectrum (\citealt{1998A&A...330.1109A}, \citealt{Plez2012}). 
The analysis was carried out differentially with respect to the Sun, whereby a line-by-line investigation was applied. 
As the target stars were observed using two resolutions ($\sim 36\,000$ and $\sim 68\,000$), 
their spectra were investigated differentially to the solar spectra observed in the corresponding resolution mode. The abundances were averaged from several observations, if available. Individual uncertainties in determining the atmospheric parameters and abundances of chemical elements are presented in \citetalias{Tautvaisiene2020}, \citetalias{Tautvaisiene2021}, and \citetalias{Tautvaisiene2022}.  

\subsubsection{Non-local thermodynamic equilibrium effects}

\begin{figure}
    \includegraphics[width=\columnwidth]{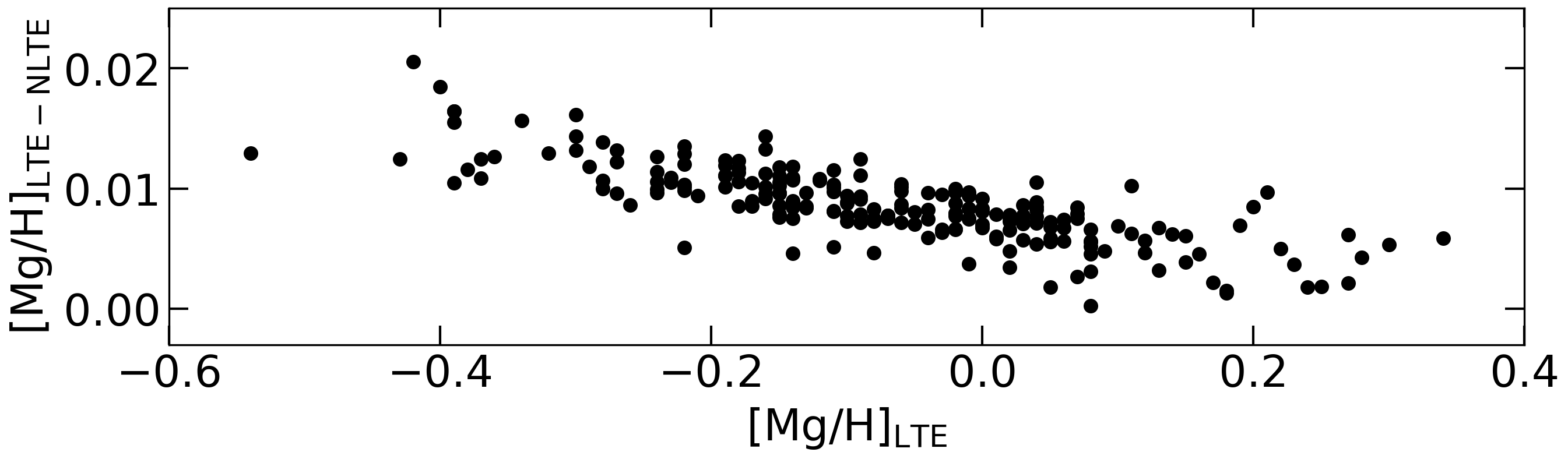}
        \includegraphics[width=\columnwidth]{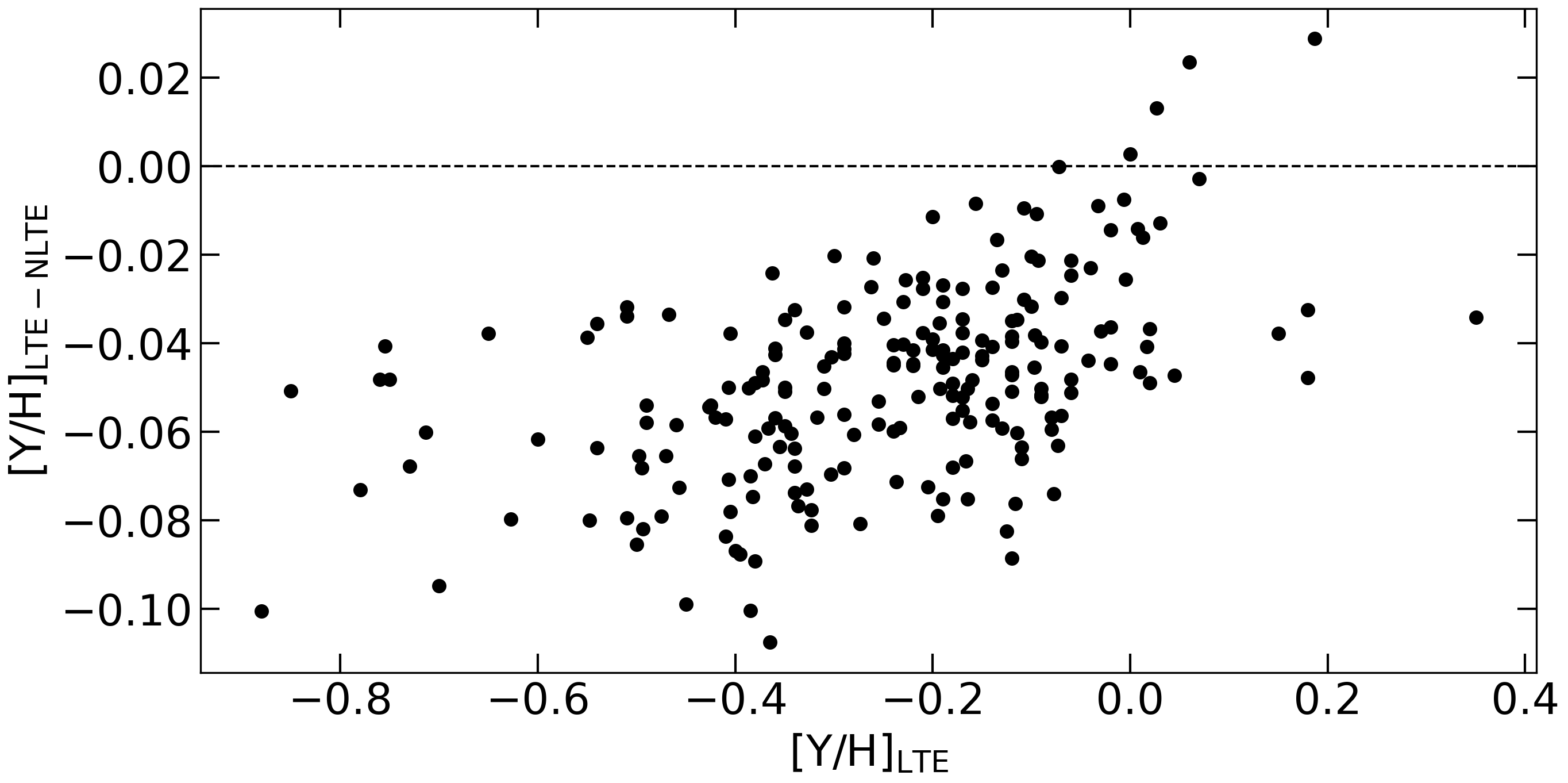}
    \caption {NLTE corrections of [Mg/H] and [Y/H] for the investigated stars. }
    \label{fig:nlte}
\end{figure}

For the robustness of the study, we paid special attention to accounting for non-local thermodynamic equilibrium (NLTE) effects.   
As we focused on [C/N] and [Y/Mg], it is worth mentioning that C$_2$ bands that were used to determine the carbon abundance are not sensitive to NLTE deviations (\citealt{Clegg1981, Gustafsson1999}). The 6300.3~\AA\ oxygen forbidden line used in the analysis is known to be unaffected by NLTE and shows little sensitivity to 3D effects (\citealt{Asplund2004, Pereira2009}). This line forms nearly in LTE and is only weakly sensitive to convection; its formation is similar in 3D radiation hydrodynamic and 3D magnetohydrodynamic solar models (\citealt{Bergemann21}).
Therefore, possible NLTE effects on [C/N] should be very small.
 However, for magnesium and yttrium abundances we decided to compute the NLTE corrections according to \cite{Storm2023} and \cite{Storm2024} for Y\,{\sc ii} and according to \cite{Bergemann2017} for Mg\,{\sc i} using the updated Turbospectrum code (\citealt{Gerber2023}).
 As in \citet{Adibekyan2017}, the NLTE effects on the [Mg/H] ratios in the  investigated metallicity interval are small; however, for [Y/H] they reach $\sim 0.1$~dex. The dependencies are illustrated in Fig.~\ref{fig:nlte}.

\subsection{Asteroseismic ages}

We analysed 30-minute and two-minute cadence TESS light curves of our set of stars observed in sectors 13--26. 
For targets observed with the 30-minute cadence, which constituted the majority of the analysed sample, the light curves were extracted from the  TESS full-frame images using the TESSCut service \citep{2019ascl.soft05007B}  using the Lightkurve package \citep{LightkurveCollaboration2018}. For targets observed with two-minute cadence, the calibrated light curves were retrieved via the TESS bulk-download service, based on products generated by the SPOC pipeline \citep{2016SPIE.9913E..3EJ}.
The light curves were de-trended and sigma clipped using a combination of standard procedures available within the Lightkurve package and custom routines developed by the authors when the default de-trending methods were found to be insufficient for asteroseismic analysis. Light curves from all available sectors were combined prior to analysis to improve the signal-to-noise ratio and frequency resolution, which is particularly important for low-frequency oscillations in red giant stars.
Power spectra were computed from the de-trended combined light curves using the standard Fourier-based methods implemented in the Lightkurve package. The resulting power spectral densities were visually inspected to assess data quality and to confirm the presence of oscillation power excess.
The frequency at maximum oscillation power ($\nu_{\rm{max}}$) was estimated by fitting a Gaussian profile to the continuum-flattened power spectrum and identifying the frequency corresponding to the peak of the fit. The large frequency separation ($\Delta\nu$) was determined using the autocorrelation of the power spectrum in the frequency region around $\nu_{\rm{max}}$ and subsequently manually verified using échelle diagrams. The uncertainty in $\nu_{\rm{max}}$ was estimated from the full width at half maximum (FWHM) of the fitted Gaussian envelope of the oscillation power excess. The uncertainty in $\Delta\nu$ was estimated from the width of the autocorrelation peak and by visual inspection of the corresponding échelle diagram.

\begin{figure}
    \centering
    \includegraphics[width=0.47\columnwidth]{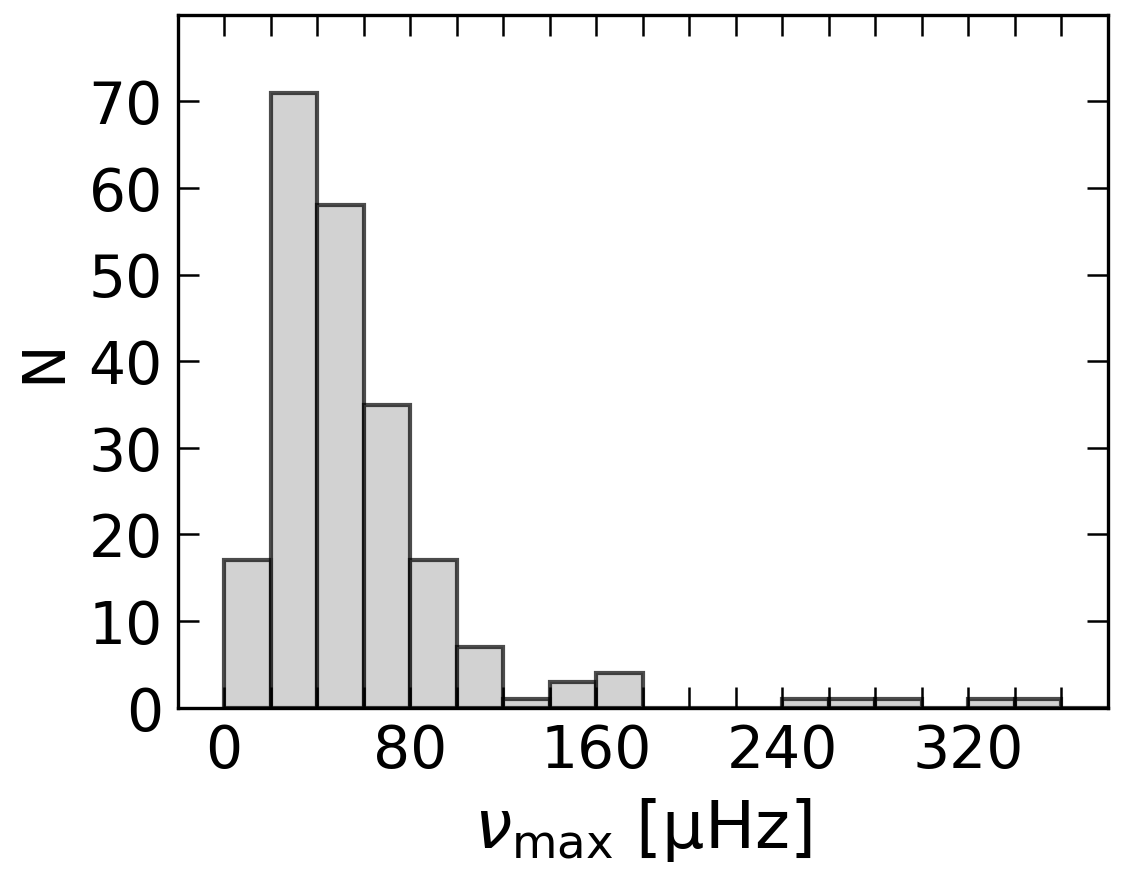}
    \includegraphics[width=0.47\columnwidth]{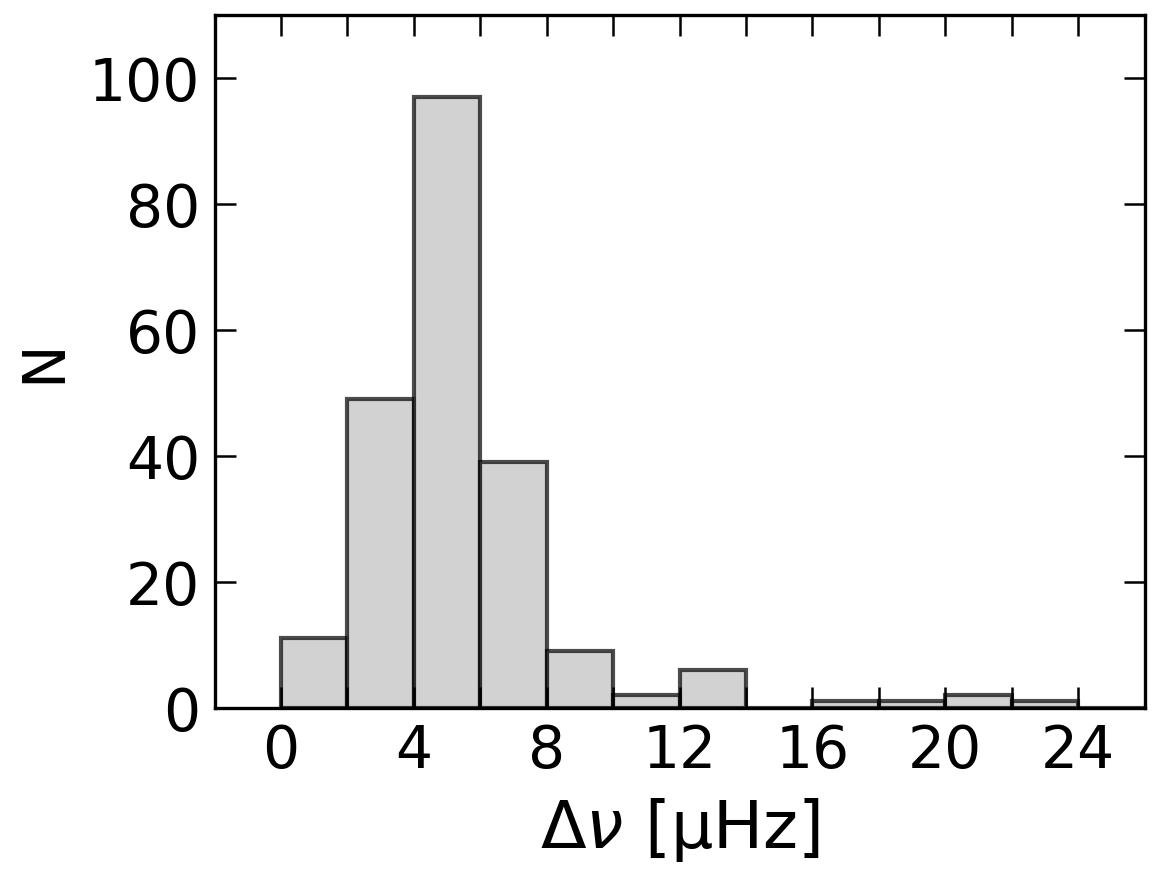}
    \caption{Distribution of stars according to frequency at maximum power, $\nu_{\rm{max}}$, and the large frequency separation, $\Delta\nu$, derived in this study.}
    \label{histo-seism}
\end{figure}

This enabled us to obtain the pulsation spectra of all our stars; however, robust determinations of $\nu_{\rm{max}}$ and $\Delta\nu$ were possible for only 218 stars. 
Only stars exhibiting clear solar-like pulsations with easily recognisable ridges of $\ell =$ 0, 1, and 2 modes in echelle diagrams were selected for further analysis and age determination. The derived asteroseismic parameters were visually checked for each star to evaluate their reliability and corrected if necessary. We were unable to determine the values of $\nu_{\rm{max}}$ and  $\Delta\nu$ for stars with $T_{\rm eff} < 4200$~K and $T_{\rm eff} > 5200$~K, and with log\,$g < 2$ and log\,$g > 3.4$. For other stars, the success rate of robust determination of asteroseismic parameters depended on the number of pointings. In the CVZ, it was about 60\%; while in the area around it, it was about 20\%.
Consequently, the number of stars with asteroseismic ages determined was lower than we expected; however, it was sufficient to achieve the aims of this study.
Figure~\ref{histo-seism} shows the distribution of stars with $\nu_{\rm{max}}$ and $\Delta\nu$ determined. 

Using $\nu_{\rm max}$, $\Delta\nu$, and the spectroscopic effective temperature $T_{\rm{eff}}$, we calculated the initial approximate stellar mass, radius, $\log g$, and luminosity based on classical scaling relations \citep{KjeldsenBedding1995}: 

\begin{equation} \label{eq1}
\frac{M}{M_\odot} \simeq \left(\frac{\nu_{\rm max}}{\nu_{\rm max,\odot}}\right)^3\left(\frac{\Delta\nu}{\Delta\nu_{\rm\odot}}\right)^{-4}\left(\frac{T_{\rm eff}}{T_{\rm eff,\odot}}\right)^{3/2},
\end{equation}

\begin{equation} \label{eq2}
\frac{R}{R_\odot} \simeq \left(\frac{\nu_{\rm max}}{\nu_{\rm max,\odot}}\right)\left(\frac{\Delta\nu}{\Delta\nu_{\rm\odot}}\right)^{-2}\left(\frac{T_{\rm eff}}{T_{\rm eff,\odot}}\right)^{1/2},   
\end{equation}

\begin{equation} \label{eq3}
\log g \simeq \log\left(g_{\odot}\left(\frac{M}{M_\odot}\right)\left(\frac{R_\odot}{R}\right)^2\right),
\end{equation}

\begin{equation} \label{eq4}
\frac{L}{L_\odot} \simeq \left(\frac{R}{R_\odot}\right)^2\left(\frac{T_{\rm eff}}{T_{\rm eff,\odot}}\right)^{4}.   
\end{equation}Then, two computing codes were used to determine the asteroseismic age and precise stellar parameters. The first was an online interface PARAM (v.1.5) for the Bayesian estimation of stellar parameters \citep{daSilva2006, Rodrigues2014, Rodrigues2017}, and the second was the BAyesian STellar algorithm (BASTA; \citealt{Aguirre2015,Aguire2022}).

\subsubsection{Age determinations with PARAM}

The parameters used for the age calculations in PARAM included $\nu_{\rm max}$, $\Delta\nu$, spectroscopic $T_{\rm{eff}}$, and metallicity [Fe/H], along with asteroseismic $\log g $ and luminosity $L/L_\odot$ (both calculated using scaling relations \ref{eq3} and \ref{eq4}, which helped to reduce age-determination uncertainties and did not materially change output values). We assumed the solar seismic parameters $\nu_{\rm max, \odot}=3141~\mu$Hz and $\Delta\nu_\odot=134.98~\mu$Hz \citep{FredslundAndersen2019}, a solar metal content of $Z_\odot=0.01756$ \citep{Paxton2011, Paxton2013}, an exponential initial mass function \citep{Chabrier2001}, and an unknown prior of the evolutionary stage. 
To account for the mass loss in our calculations, we applied a Reimers-type mass-loss efficiency coefficient, $\eta_R = 0.4$. This value reflects the average value of the Reimers coefficient well considering different empirical contexts and models \citep{McDonald&Zijlstra2015, Valle&DellOmodarme2018}.

\begin{figure}
    \centering
    \includegraphics[width=0.47\columnwidth]{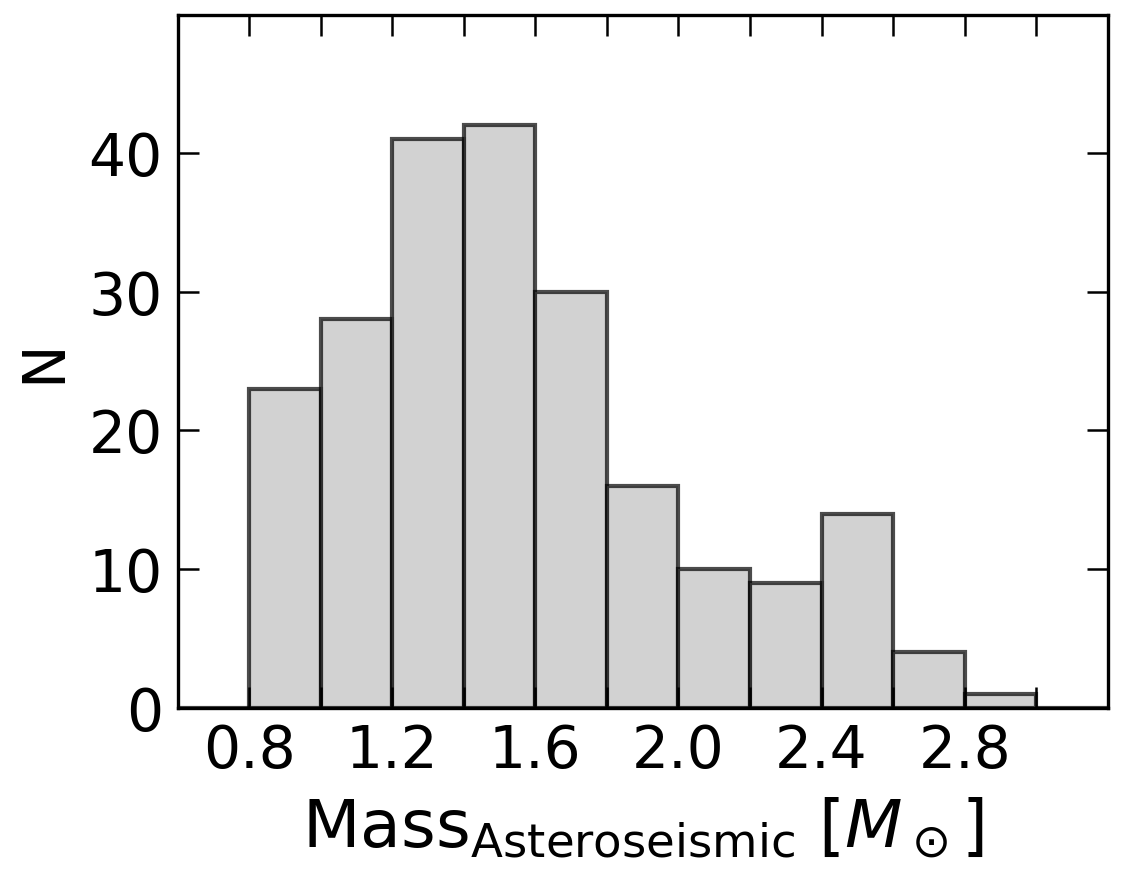}
    \includegraphics[width=0.47\columnwidth]{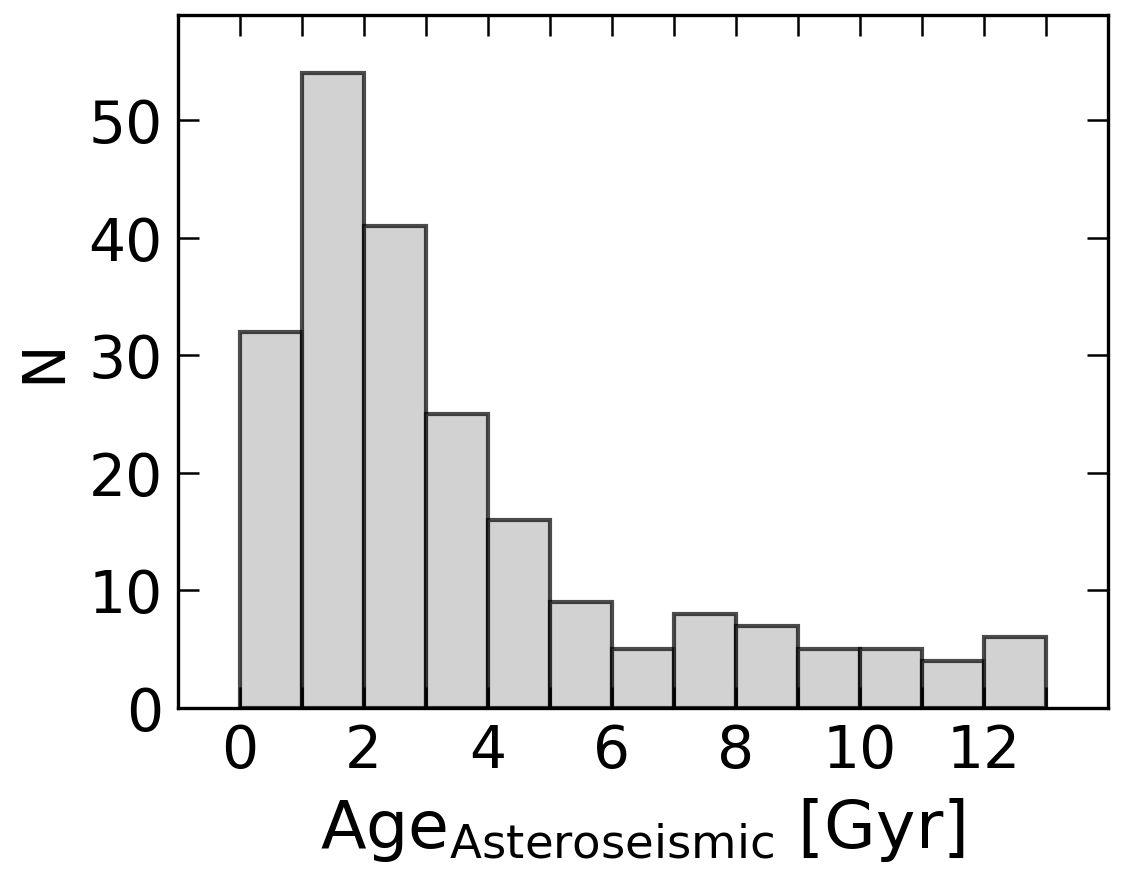}   
    \caption{Distribution of stars according to their asteroseismic masses and ages  determined in this study.}
    \label{histo-mass-age}
\end{figure}

The ages were determined using a grid of MESA isochrones \citep{Rodrigues2017}, where individual radial mode frequencies and large separations, $\Delta\nu,$ were calculated for each model of the grid. As the PARAM description states that MESA isochrones are available in the mass range of $0.6 \leq M/M_\odot < 2.5$, 
for stars with a scaling relation mass greater than $2.5 M_\odot$ (up to $4\,M_\odot$), the age was calculated with the PARSEC isochrones instead. Finally, a two-step Bayesian method \citep{Rodrigues2014} was used to determine the ages of our selected stars. Using this method, the ages were determined for 218 stars. 

\subsubsection{Age determinations with BASTA}

This algorithm uses grid-based modelling to compare the observed parameters of the stars with stellar models and to infer the parameters of the stars using Bayesian statistics.
Here, the observables used were the spectroscopic effective temperature, $T_{\rm{eff}}$; metallicity, $[\rm{Fe}/\rm{H}]$; frequency of maximum power, $\nu_{\rm max}$; and large frequency separation, $\Delta\nu,$ determined in this study.

The observations were compared to the BaSTI isochrones \citep{Hidalgo2018,Pietrinferni2021}, specifically case 4 in Table~1 of \cite{Aguire2022}, where the effects of overshooting, diffusion, and mass-loss are included in the models. 
Here, the model-predicted value of $\nu_{\rm max}$  
was determined using the classical scaling relations \citep[Eq. 10,][]{KjeldsenBedding1995}, and the  model-predicted value of $\Delta\nu$ was calculated using the corrected scaling relations of \cite{Sharma2016} and \cite{Stello2022}.
Using this method, the ages were determined for 182 stars, given as the median, 16th, and 84th quantiles of the posterior age distribution, as listed in Table~\ref{table:Results} (Col 4-6).

Figure~\ref{histo-mass-age} shows the mass and age distributions of the 218 stars investigated based on asteroseismic data from the TESS telescope. The masses and ages determined using PARAM and BASTA software were averaged. 

\subsection{Evolutionary stages}

For the use of C and N for the stellar age evaluation, we had to define evolutionary stages of stars (see \citealt{Tautvaisiene2025}). With this purpose in mind, asteroseismology helps significantly. 

At later stages of stellar evolution, such as the red giant branch (RGB) and helium-core-burning phases --the red clump (RC) in the case of our sample-- $g$ modes originating in the deep stellar interior begin to influence surface $p$ modes through mode coupling, creating detectable mixed modes (\citealt{Beck2011}). 
 This coupling leads to characteristic period spacings ($\Delta P$) between the consecutive mixed $\ell =1,$  reflecting the underlying $g$-mode structure. This period spacing, also called the bumped period spacing (e.g. \citealt{Mosser2015}), is particularly sensitive to the core structure, allowing the period spacing to reflect differences in burning processes, such as hydrogen-shell burning in RGB versus the core helium burning in RC stars.  
Studies revealed that the period spacings between consecutive $\ell =1$ mixed modes are directly sensitive to the density gradient (or Brunt–Väisälä frequency profile) in the radiative zone between the core and convective envelope (\citealt{Beck2011}; \citealt{Beddingetal2011}; \citealt{Mosser2012}). The interpretation of period spacing as a function of stellar oscillations has been confirmed and refined in subsequent studies (e.g. \citealt{Mosser2014}; \citealt{Mosser2018}).

In stars undergoing hydrogen shell burning on the RGB, the buoyancy frequency structure leads to relatively small period spacing ($\Delta P \approx 30-80$~s). In contrast, $\Delta P$ in stars undergoing helium burning in the core on the horizontal branch or in the red clump is larger ($\Delta P \gtrsim 100$~s) due to differences in the core structure and surrounding stratification. This distinction provides a robust way to identify the evolutionary stage of red giants (\citealt{Beddingetal2011}). 

\begin{figure}
    \centering
    \includegraphics[width=1.0\columnwidth]{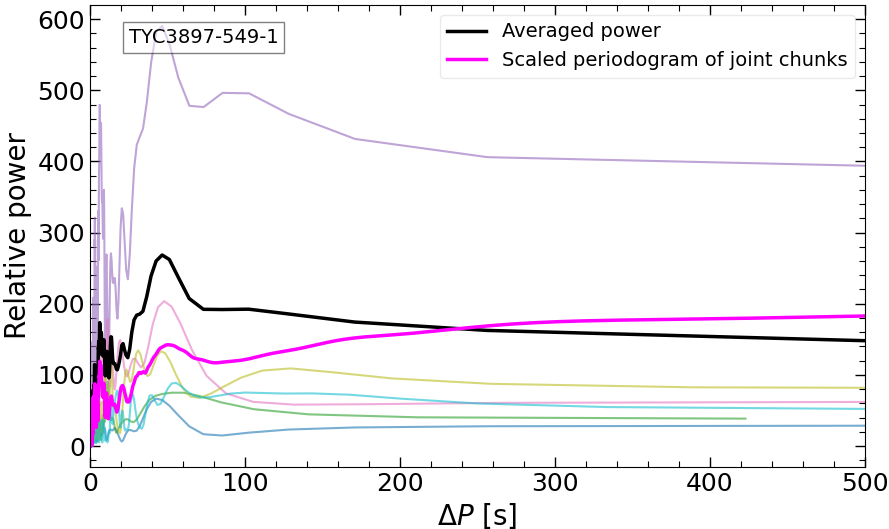}
    \includegraphics[width=1.0\columnwidth]{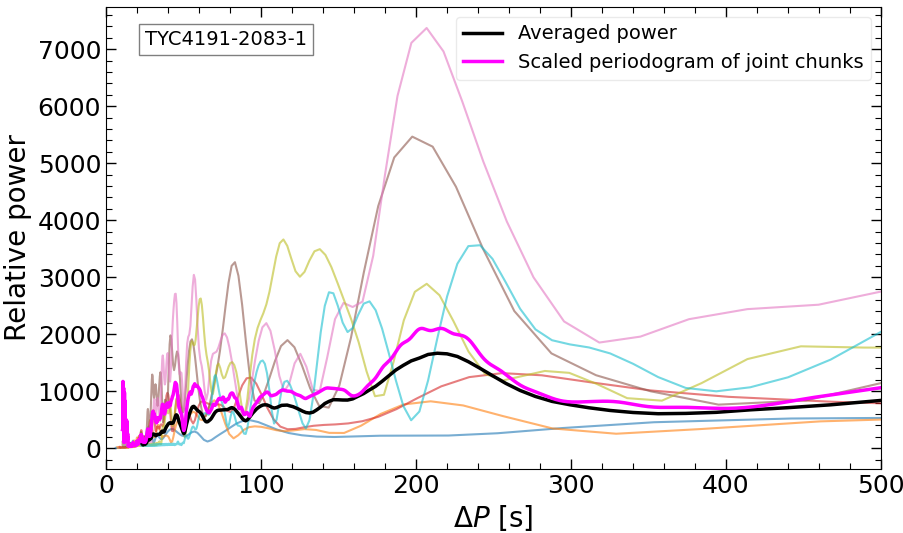}
    \caption{Examples of 'periodograms of periodograms' for the RGB star TYC3897-549-1 and the RC star TYC4191-2083-1. }
    \label{examples}
\end{figure}

\begin{figure}
    \centering
    \includegraphics[width=1.0\columnwidth]{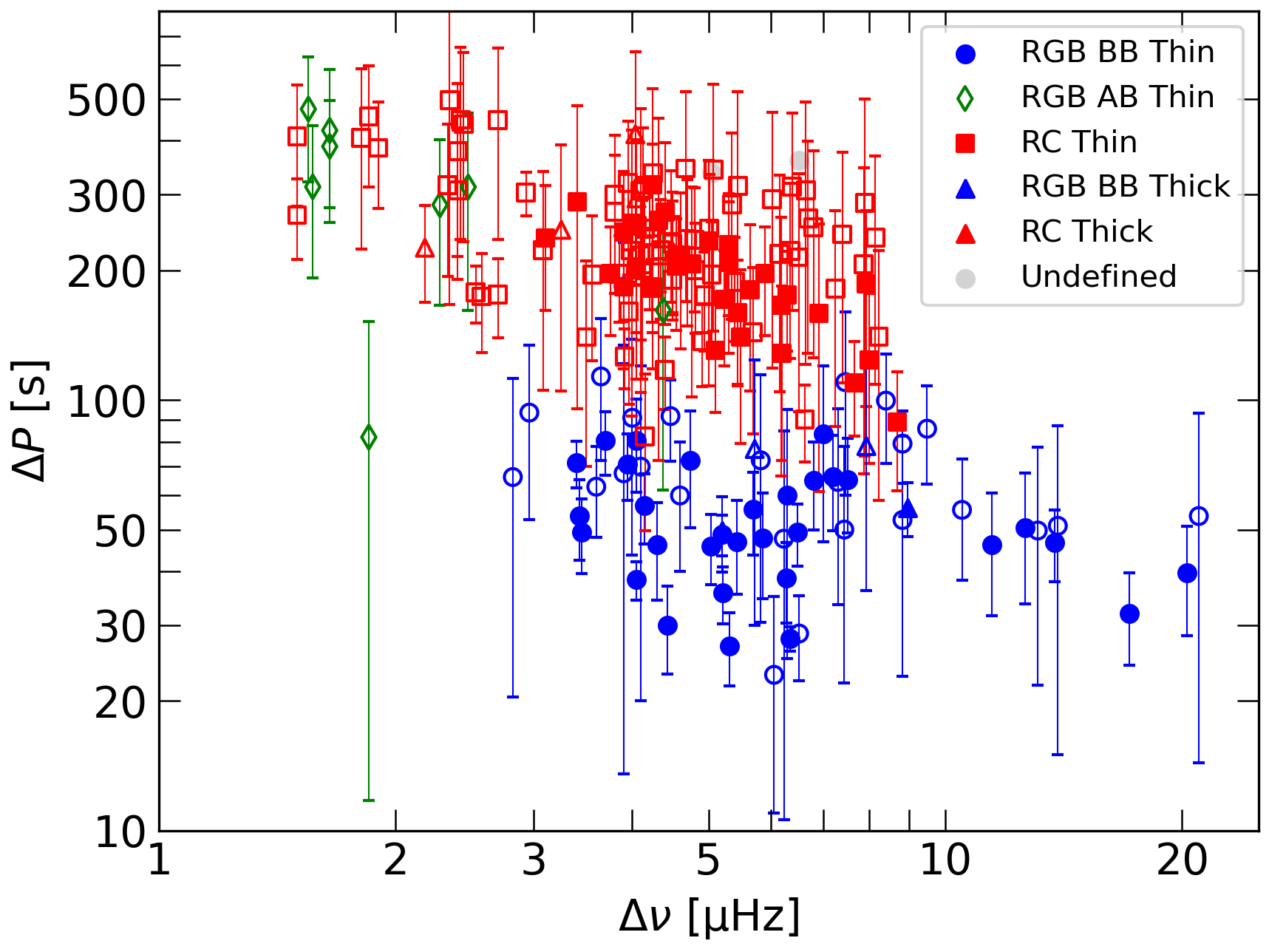}
    \caption{ Relation of $\Delta P$ and $\Delta \nu$ for the investigated stars. The blue symbols show giants below the RGB luminosity bump (BB), the green symbols are for giants above the bump (AB), and the red symbols are for red clump (RC) stars. The filled circles show stars for which the evolutionary stage was attributed with higher confidence. See the text for more explanations. } 
    \label{fig:DP_Dnu}
\end{figure}

Thus, we analysed the $\ell =1$ pulsation mode regions in the pulsation periodograms to determine the most likely $\Delta P$ value. For this purpose, we computed a periodogram for each selected spectral chunk containing split $\ell =1$ modes (pale coloured curves in Fig.~\ref{examples}), as well as their averaged periodogram of periodograms, and that obtained from the combined chunks (see Fig.~\ref{examples}). 
Period spacing values were obtained for 177 stars, and the distribution of the resulting $\Delta P$ and $\Delta \nu$ is shown in Fig.~\ref{fig:DP_Dnu}. The results of higher confidence are marked with filled symbols. They represent 33 stars, for which the evolutionary stage was classified as giants below the RGB luminosity bump (RGB BB), and 32 stars were classified as RCs. For the remaining stars with $\Delta P$ of lower accuracy or without its determination, the evolutionary phases were inferred by involving investigations of stellar locations in the log\,$g$ versus $T_{\rm eff}$ diagrams relatively to the corresponding evolutionary sequences. These stars are presented in the figures by the open symbols. From the evolutionary sequences, we also inferred nine stars in the evolutionary phase above the RGB luminosity bump (RGB AB). They are marked in the figures by the green diamonds. For 20 stars, the evolutionary stage remained undefined. They are marked in the figures by the grey symbols. The triangles show stars belonging to the thick disc.  

It is interesting to note that among the 189 bright stars investigated in the TESS continuous viewing zone in our work, about 60\% belong to the RC and 40\% to the RGB. Very similar  percentages of stars in these two evolutionary stages were identified among 395 fainter stars in the $Kepler$ field by \cite{Beddingetal2011}.

\begin{figure*}
    \centering
    \includegraphics[width=0.33\textwidth]{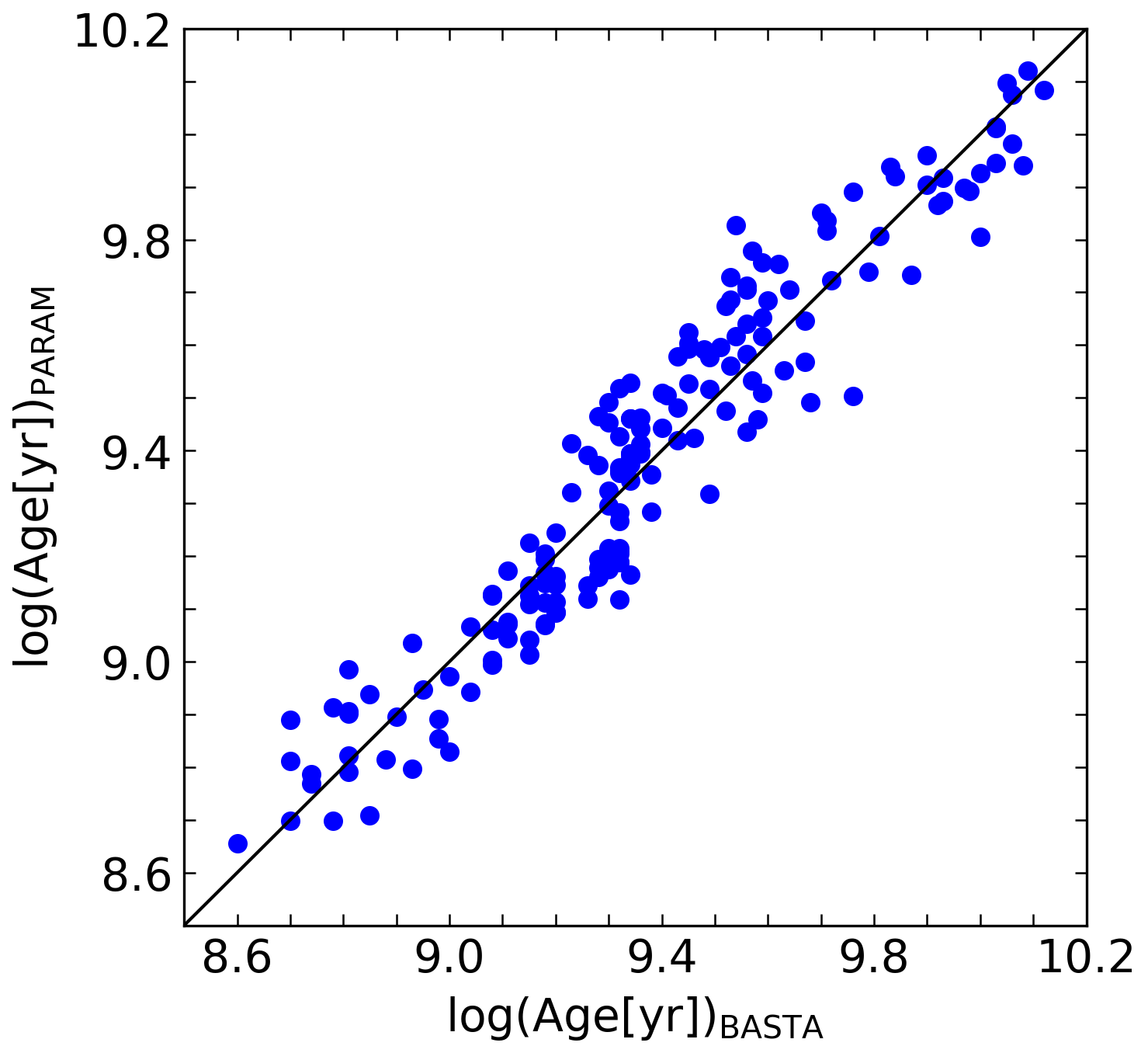}
    \includegraphics[width=0.33\textwidth]{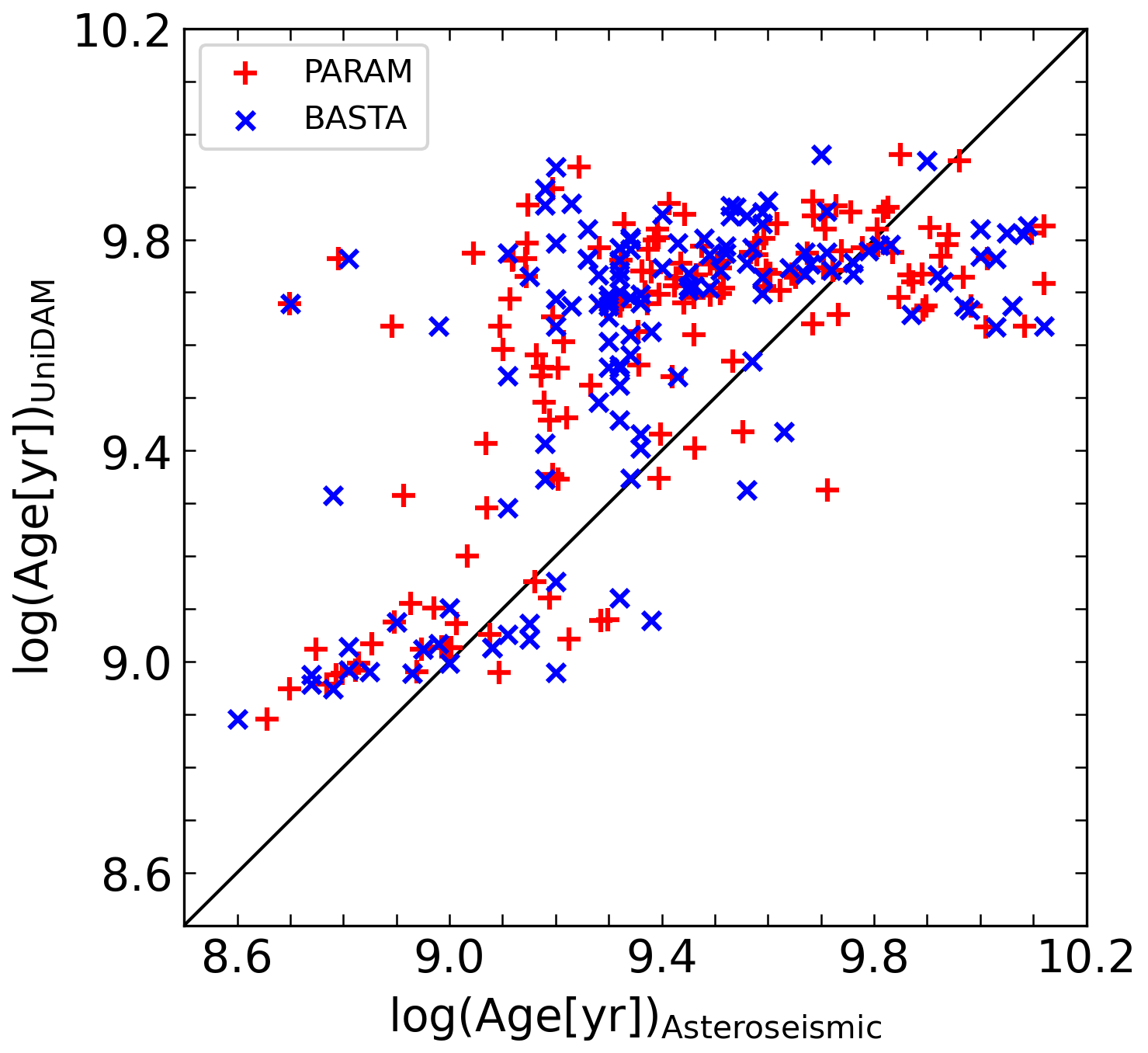}
    \includegraphics[width=0.305\textwidth]{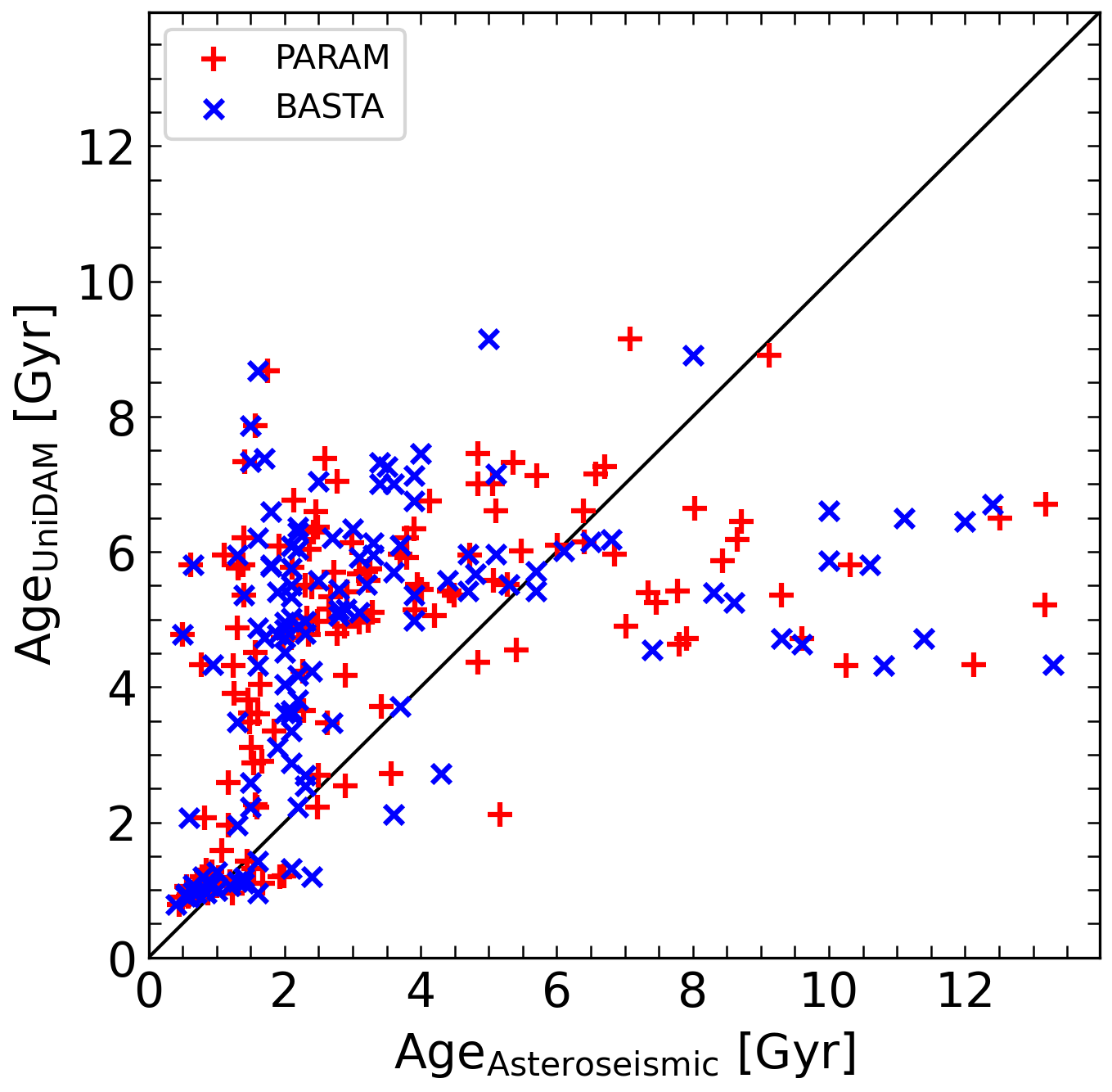}
    \caption{Comparison of ages determined using PARAM (v.1.5), BASTA, and UniDAM softwares.}
    \label{comp-age}
\end{figure*}

\section{Results and discussion}
\label{results}

In the machine-readable Table~\ref{table:Results}, we provide the determined asteroseismic stellar ages, masses, $\nu_{\rm{max}}$ and $\Delta\nu$ values, NLTE [Mg/H] and [Y/H] abundances, the determined stellar evolutionary stages, and other parameters used in this study taken from \citetalias{Tautvaisiene2020}, \citetalias{Tautvaisiene2021}, and \citetalias{Tautvaisiene2022} for convenience. 

\subsection{Comparison of asteroseismic and classical isochronal ages}

The age determination of field giants is an extremely difficult task when using the traditional technique of matching the atmospheric parameters to stellar isochrones, as the isochrones with different parameters are located quite close to each other, or even overlap. To overcome this problem, additional observables to the effective temperature, surface gravity, and metallicity have been employed and tested with a statistical Bayesian approach (e.g. \citealt{Jorgensen2005}, \citealt{2017A&A...604A.108M}). 
One such method, the UniDAM tool, uses spectroscopic data combined with infrared photometry and compares these in a Bayesian manner with PARSEC isochrones to derive probability density functions (PDFs) for stellar masses, ages, and distances. It treats PDFs of pre-helium-core-burning, helium-core-burning, and post-helium-core-burning solutions, as well as different peaks in multi-modal PDFs of the different stellar evolutionary phases separately (\citealt{2017A&A...604A.108M}). Using this tool, the ages were determined for all our sample of stars in  \citetalias{Tautvaisiene2020} and \citetalias{Tautvaisiene2022}. We decided to compare the ages determined using the UniDAM tool with the ages determined in this work using two approaches, which involve very important asteroseismic observables.    

In Fig.~\ref{comp-age}, we first compare the ages determined with the PARAM and BASTA tools (left panel) and see a rather good agreement. The middle panel shows the comparison of UniDAM ages with PARAM and BASTA ages on a logarithmic scale, and the right panel is presented on a
linear scale for convenience. It is visible that with the UniDAM tool, the ages of young stars were overestimated, whereas for older stars, they were underestimated. Bearing in mind that solar-type oscillations are not detected in all giant stats, the classical methods have to be further developed, and maybe some new probability density functions could be adopted from asteroseismology to remove deviations. Other stellar age-determination methods, such as that of chemical clocks, investigated in this work, have also to be developed. 

\subsection{Relation of [Y/Mg] with age}

To investigate possible radial dependencies in the $[\mathrm{Y}/\mathrm{Mg}]_{\mathrm{NLTE}}$ and age relationship, we divided the sample into three subsamples according to the mean galactocentric radius of each star ($R_{\mathrm{mean}}$). Stars with $R_{\mathrm{mean}} < 7.5$~kpc were assigned to the inner disc, those with $7.5 \le R_{\mathrm{mean}} \le 8.5$~kpc define the solar neighbourhood, and stars with $R_{\mathrm{mean}} > 8.5$~kpc trace the outer disc, following commonly adopted radial divisions in the literature \citep[e.g.][]{Boeche2014,Hunt2015, Sysoliatina2018}. 

\begin{figure}
    \centering
    \includegraphics[width=1.0\columnwidth]{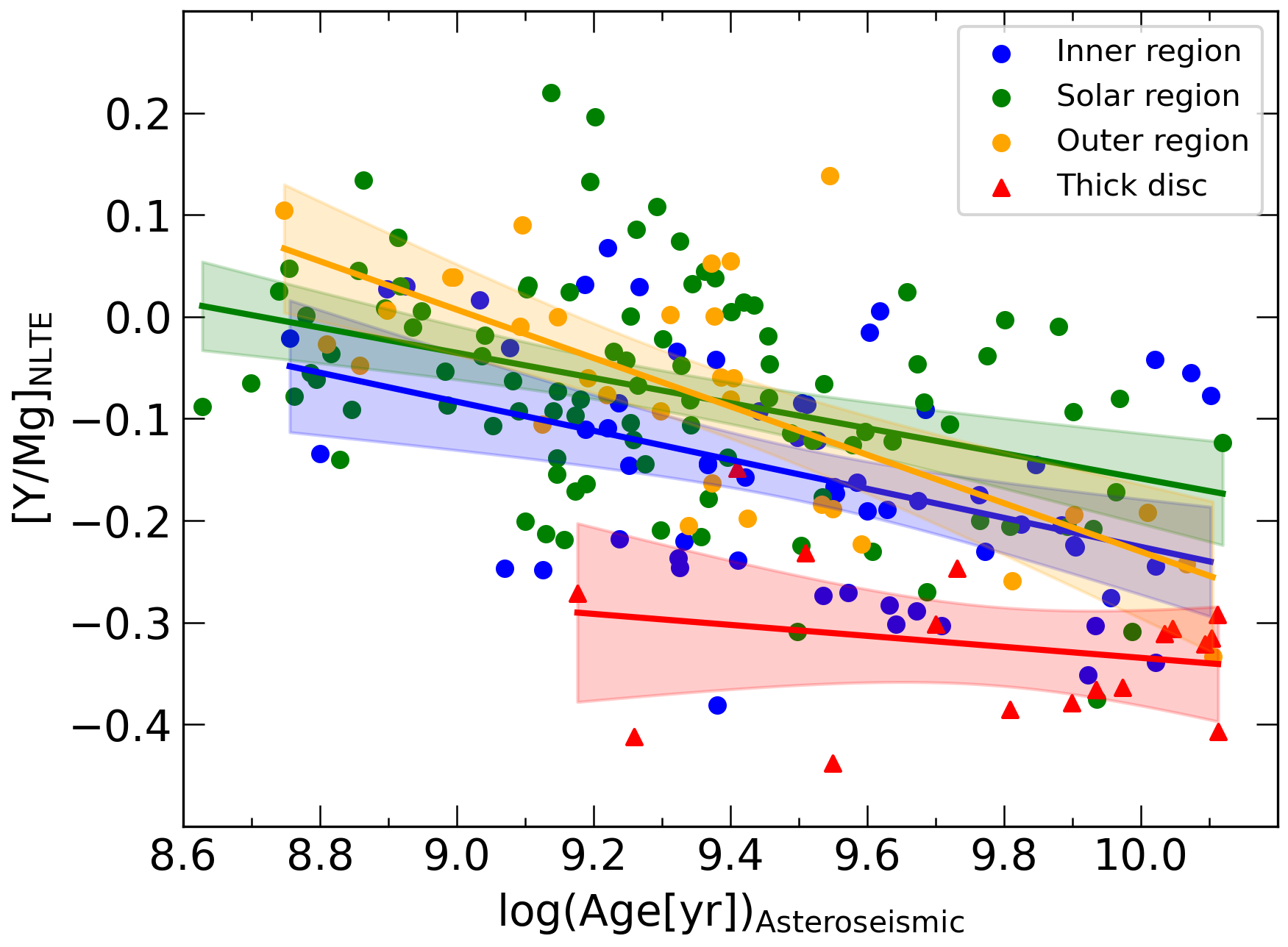}
    \caption{[Y/Mg] as a function of stellar age for thin- and thick-disc stars. Linear fits for different Galactic regions and their 95\% confidence intervals are shown. Stars with $R_{\rm mean} < 7.5$~kpc are attributed to the inner Galactic disc, the solar region is at $7.5~\leq~R_{\rm mean}~\leq~8.5$~kpc, and the outer disc is at $R_{\rm mean} > 8.5$~kpc.}
    \label{YMg-age-common}
\end{figure}

Thick-disc stars were treated as a separate population based on their chemical and kinematic properties. The separation into Galactic thin- and thick-disc components of our sample stars  was done in \citetalias{Tautvaisiene2020} and \citetalias{Tautvaisiene2022} following the chemical and kinematic criteria defined in the Tinsley--Wallerstein and Toomre diagrams, as well as the age-chemo-kinematics approach proposed by \cite{Lagarde2021}.
This division allowed us to explore how the $[\mathrm{Y}/\mathrm{Mg}]_{\mathrm{NLTE}}$--age relation varies across the Galaxy, and the resulting trends are shown together with their linear fits in Fig.~\ref{YMg-age-common}.

A common feature of $s$-process chemical clocks is that their behaviour varies across the Galactic disc, reflecting differences in star-formation efficiencies and enrichment timescales. In our sample, this diversity is clearly visible in the $[\mathrm{Y}/\mathrm{Mg}]_{\mathrm{NLTE}}$--age relations illustrated in Fig.~\ref{YMg-age-common} and quantified by the linear relations and Pearson correlation coefficients (PCC) for all four Galactic components (Eqs.~\ref{eq:inner_logAge_YMg} -- \ref{eq:thick_logAge_YMg}): 

\begin{itemize}

    \item
Inner disc (61 stars)\\
\vspace{-0.22in}
\begin{align}
{\rm \log\,Age = -7.042\,[Y/Mg]}_{\mathrm{NLTE}} + 8.444, {\rm PCC=-0.42},
\label{eq:inner_logAge_YMg}
\end{align}

\item
Solar region (98 stars)\\
\vspace{-0.22in}
\begin{align}
{\rm \log\,Age = -8.13\,[Y/Mg]}_{\mathrm{NLTE}} + 8.74, {\rm PCC=-0.40},
\label{eq:solar_logAge_YMg}
\end{align}

\item
Outer disc (32 stars)\\
\vspace{-0.22in}
\begin{align}
{\rm \log\,Age = -4.219\,[Y/Mg]}_{\mathrm{NLTE}} + 9.046, {\rm PCC=-0.70},
\label{eq:outer_logAge_YMg}
\end{align}

\item
Thick disc (17 stars)\\
\vspace{-0.22in}
\begin{align}
{\rm \log\,Age = -18.519\,[Y/Mg]}_{\mathrm{NLTE}} + 3.778, {\rm PCC=-0.23.}
\label{eq:thick_logAge_YMg}
\end{align}

\end{itemize}

The outer disc shows the steepest decline with age (see Eq. \ref{eq:outer_logAge_YMg}), pointing to a strong temporal evolution of the $s$-process contribution driven by slow star formation and extended enrichment histories. In addition to this steeper slope, the outer disc is systematically offset towards higher [Y/Mg] values at a fixed age, indicating a higher overall level of chemical enrichment. In contrast, the inner disc and the solar neighbourhood (Eqs. \ref{eq:inner_logAge_YMg} and \ref{eq:solar_logAge_YMg}) exhibit noticeably shallower trends and lower zero points, suggesting that rapid early evolution in these regions results in a reduced sensitivity of $[\mathrm{Y}/\mathrm{Mg}]$ to stellar age. This behaviour is consistent with previous findings showing weaker [$s$-/$\alpha$]--age relations and lower zero points at smaller galactocentric radii \citep{Vazquez22, Ratcliffe24}, based on open clusters from the Gaia-ESO Survey \citep{Randich2022} and on large samples of disc field stars from APOGEE \citep{Majewski2017} and GALAH \citep{DeSilva2015}, respectively. However, the latest multi-zone chemical-evolution models, while able to reproduce the increase of [$s$-/$\alpha$] with age in the outer regions, still fail to match the observed [Y/Mg] trends in the inner disc towards young ages \citep{Molero25}. These models are based on the three-infall chemical-evolution framework with state-of-the-art nucleosynthesis prescriptions developed by \citet{Spitoni2023} and extended to the entire Galactic disc by \citet{Palla2024}. From the stellar-evolution perspective, magnetic-buoyancy-induced mixing in asymptotic giant branch (AGB) stars may reduce yttrium production at high metallicity \citep{Magrini2021}, potentially explaining the low [Y/Mg] ratios in the inner disc.

\begin{figure}
    \centering
    \includegraphics[width=1.0\columnwidth]{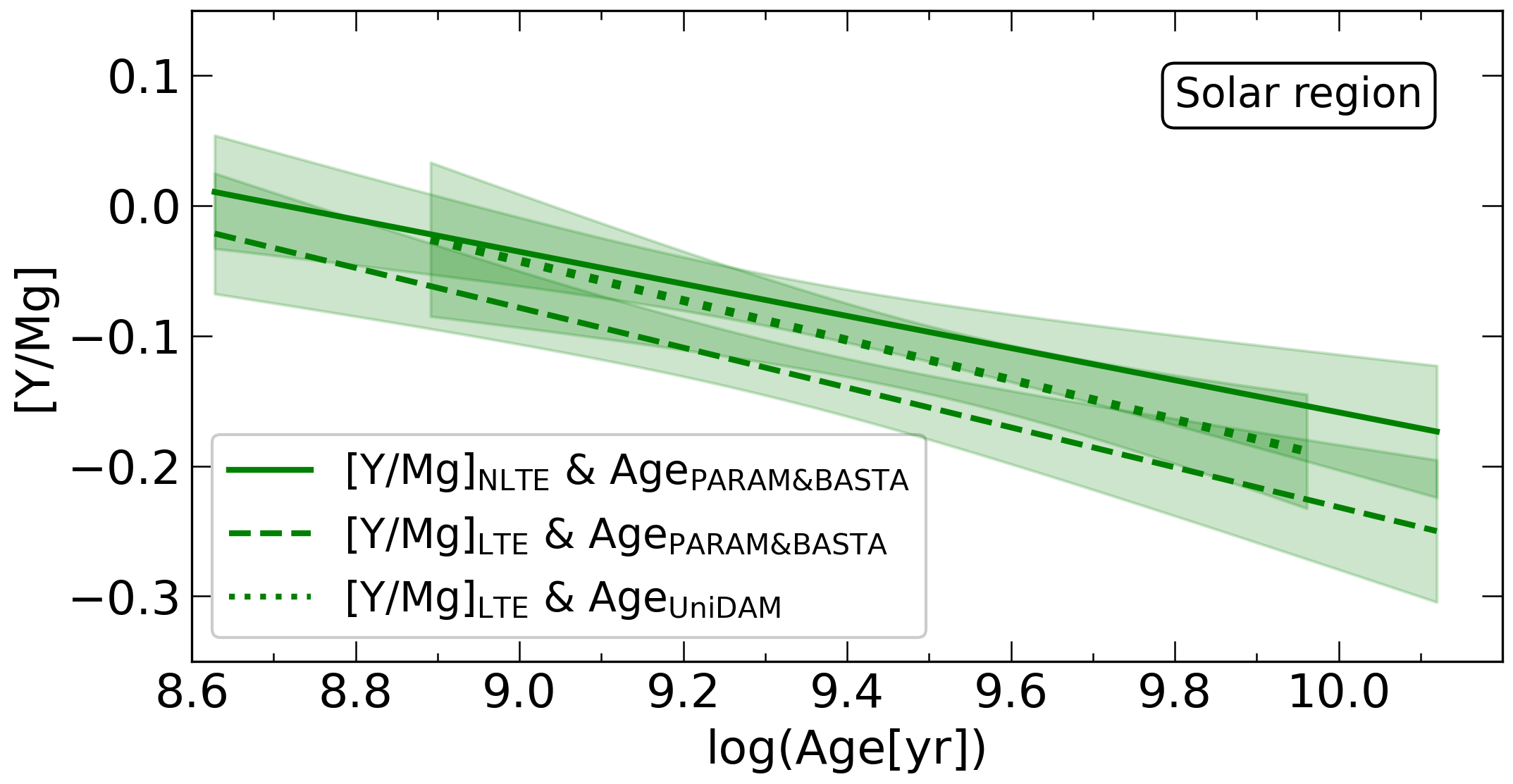}
    \caption{Comparison of [Y/Mg]$_{\rm NLTE}$ and [Y/Mg]$_{\rm LTE}$ relations with ages determined using asteroseismic data (the continuous and dashed lines, respectively) and using the LTE [Y/Mg] values and ages computed with the UniDAM software (the dotted line). The relations are for the same 99 stars in the solar region at $7.5 \leq R_{\rm mean} \leq 8.5$~kpc. The dashed relation was taken from \cite{galaxies13060136}.}
    \label{Y/Mg-comp}
\end{figure}
The thick disc, however, displays an extremely weak dependence of $[\mathrm{Y}/\mathrm{Mg}]_{\mathrm{NLTE}}$ on age and the lowest zero point (see Eq. \ref{eq:thick_logAge_YMg}). The flat trend reflects its formation during a short and intense star-formation episode dominated by Type~II supernovae, before low-mass AGB stars contributed significantly to the $s$-process enrichment, and the low zero point reflects its early formation. As a consequence, thick-disc stars retain uniformly low $[\mathrm{Y}/\mathrm{Mg}]$ values and do not show a clear age signature, confirming that this ratio is not a reliable age indicator for this older population \citepalias{Tautvaisiene2021}.
We also divided stars of the thin disc according to their maximum distance from the Galactic plane into three groups with |$z_{\rm max}$| from 0 to 0.2~kpc, from 0.2 to 0.4~kpc, and with distances  larger than 0.4~kpc, but a negligible difference in the relations of [Y/Mg] with age was found. 

With the aim of seeing how important it is to use the NLTE [Y/Mg] values and asteroseismic ages for the chemical clock's calibration, we compared [Y/Mg]$_{\rm NLTE}$ and [Y/Mg]$_{\rm LTE}$ relations with ages determined using asteroseismic data in this work and using the UniDAM software in \citetalias{Tautvaisiene2020} and \citetalias{Tautvaisiene2022}. In Fig.~\ref{Y/Mg-comp}, we show the comparison for 99 stars in the solar region ($7.5 \leq R_{\rm mean} \leq 8.5$~kpc). We can see systematic differences, which can lead to quite significant uncertainties in age determination. It is interesting that the relationship between [Y/Mg]$_{\rm{LTE}}$ and UniDAM ages lies between NLTE and LTE [Y/Mg] and asteroseismic age relations.      

\subsection{Relation of [C/N] with age}

\begin{figure}
    \centering
    \includegraphics[width=1.0\columnwidth]{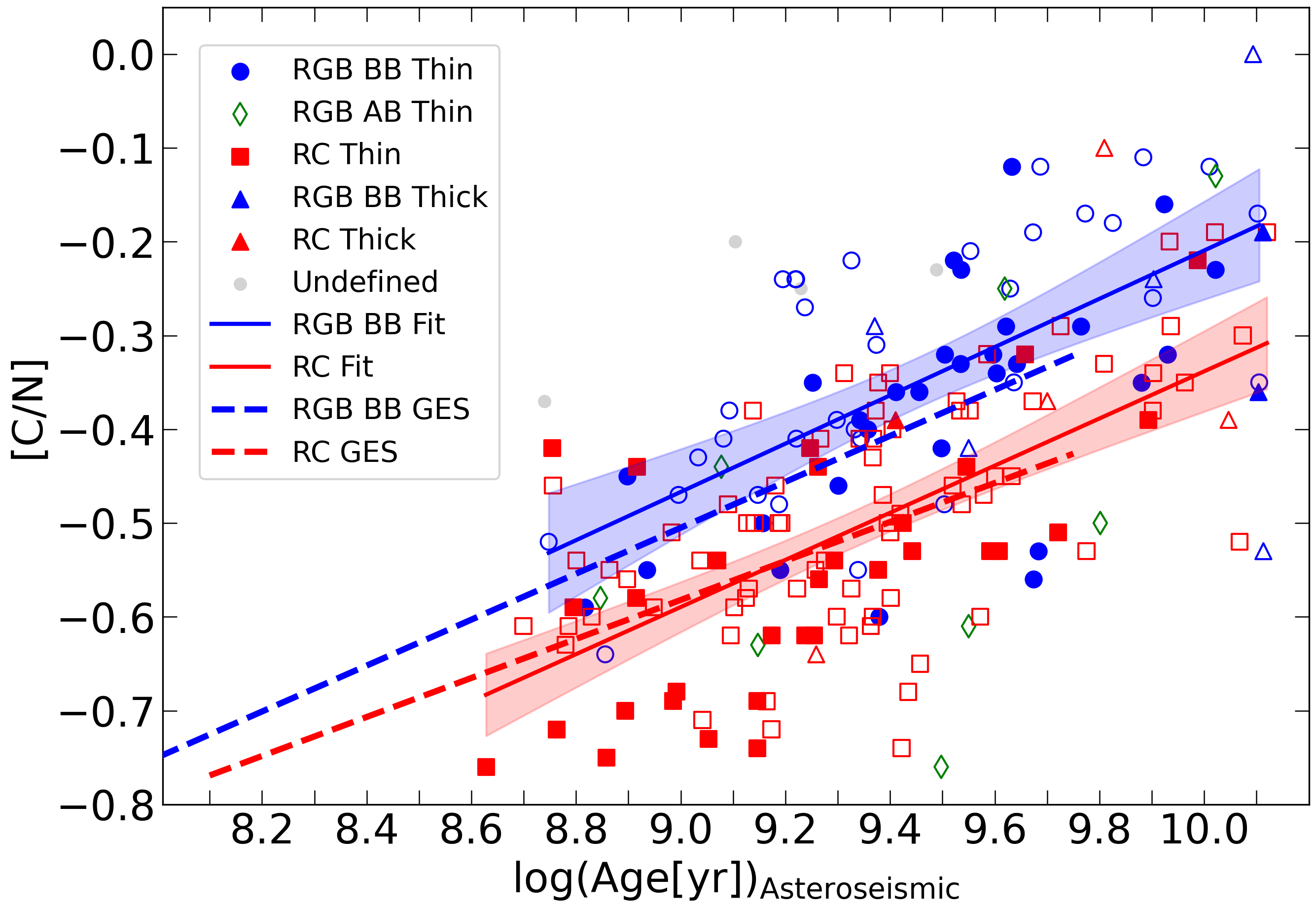}
    \includegraphics[width=1.0\columnwidth]{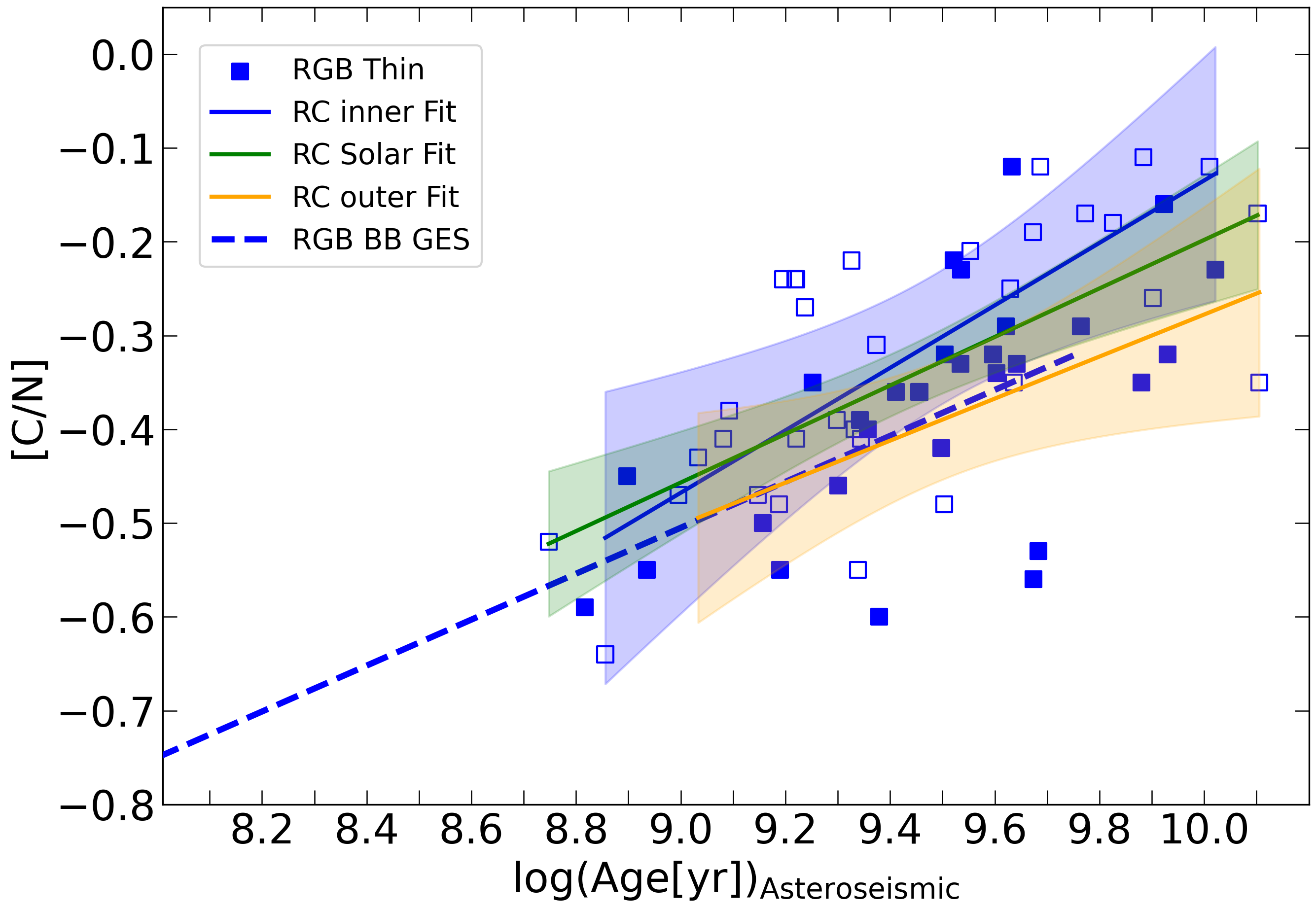}
    \includegraphics[width=1.0\columnwidth]{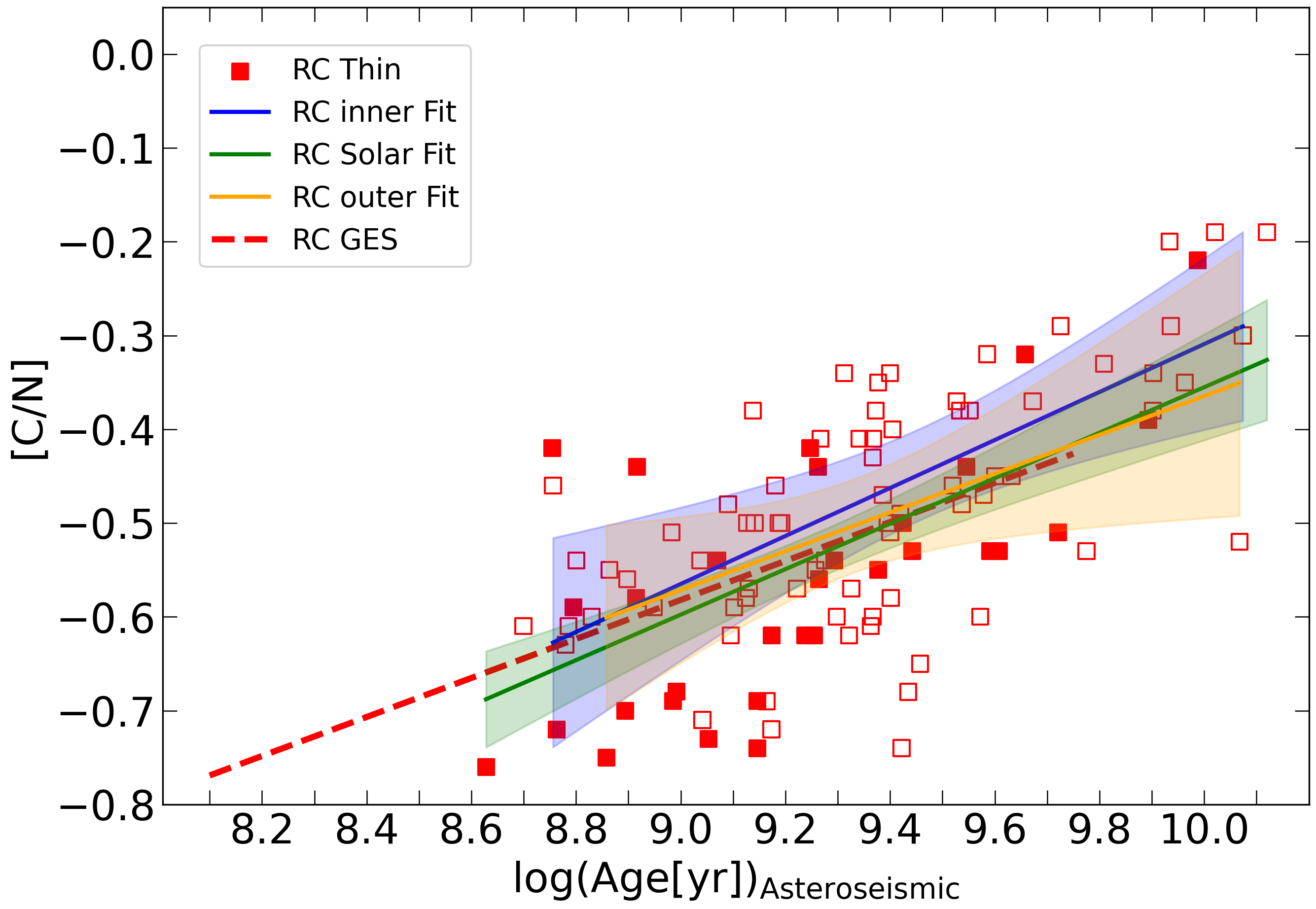}
    \caption{Relation of [C/N] with asteroseismic age for stars at different evolutionary stages.  The filled symbols are for stars with evolutionary stages determined with higher confidence. In the first plot, the solid blue relation is for the first ascent RGB stars, the red one is for RC stars, both are of the thin-disc stars. The shadowed areas show the confidence intervals of 95\%. The corresponding relations from the $Gaia$-ESO Survey \citep{Tautvaisiene2025} are shown for comparison by dashed lines. In the following two plots, the relations are shown for RGB and RC stars in the inner Galactic disc ($R_{\rm mean} < 7.5$~kpc), in the Solar region ($7.5 \leq R_{\rm mean} \leq 8.5$~kpc), and in the outer disc ($R_{\rm mean} > 8.5$~kpc), respectively.}
    \label{CN-age-common}
\end{figure}

Figure~\ref{CN-age-common} shows [C/N] dependence on stellar ages for stars at the lower part of the RGB and for helium-core-burning stars. Smaller subsamples, including stars of the upper part of RGB or the thick disc stars, are plotted but not used for calculations of the relations. The obtained [C/N] and age relations as presented in Fig.~\ref{CN-age-common}, the PCC coefficients, and the numbers of stars used for the calculations are the following:

\begin{itemize}

    \item
RGB stars below the RGB luminosity bump: \\ 
\vspace{-0.22in}
\begin{align}
\rm{log\,Age = 3.891\,[C/N] + 10.833, PPC=0.63, N=62}, 
\end{align}

\item
RC stars: \\
\vspace{-0.22in}
\begin{align}
{\rm log\,Age = 3.984\,[C/N] + 11.363, PPC=0.66, N=105.} 
\end{align}
\end{itemize}

In Fig.~\ref{CN-age-common}, we also plot the ralations taken from \cite{Tautvaisiene2025}, which were determined using data from the $Gaia$-ESO survey for the first ascent giants below the RGB luminosity bump and for the helium-core-burning clump stars in 44 open clusters. Despite the lower accuracy of the [C/N] determinations in the field stars and possible misattribution to evolutionary stages, the agreement of the relations is very good. The relations determined using the open clusters cover younger ages,  while our sample of field stars extend the relations to older ages well.

\begin{figure}
    \centering
    \includegraphics[width=1.0\columnwidth]{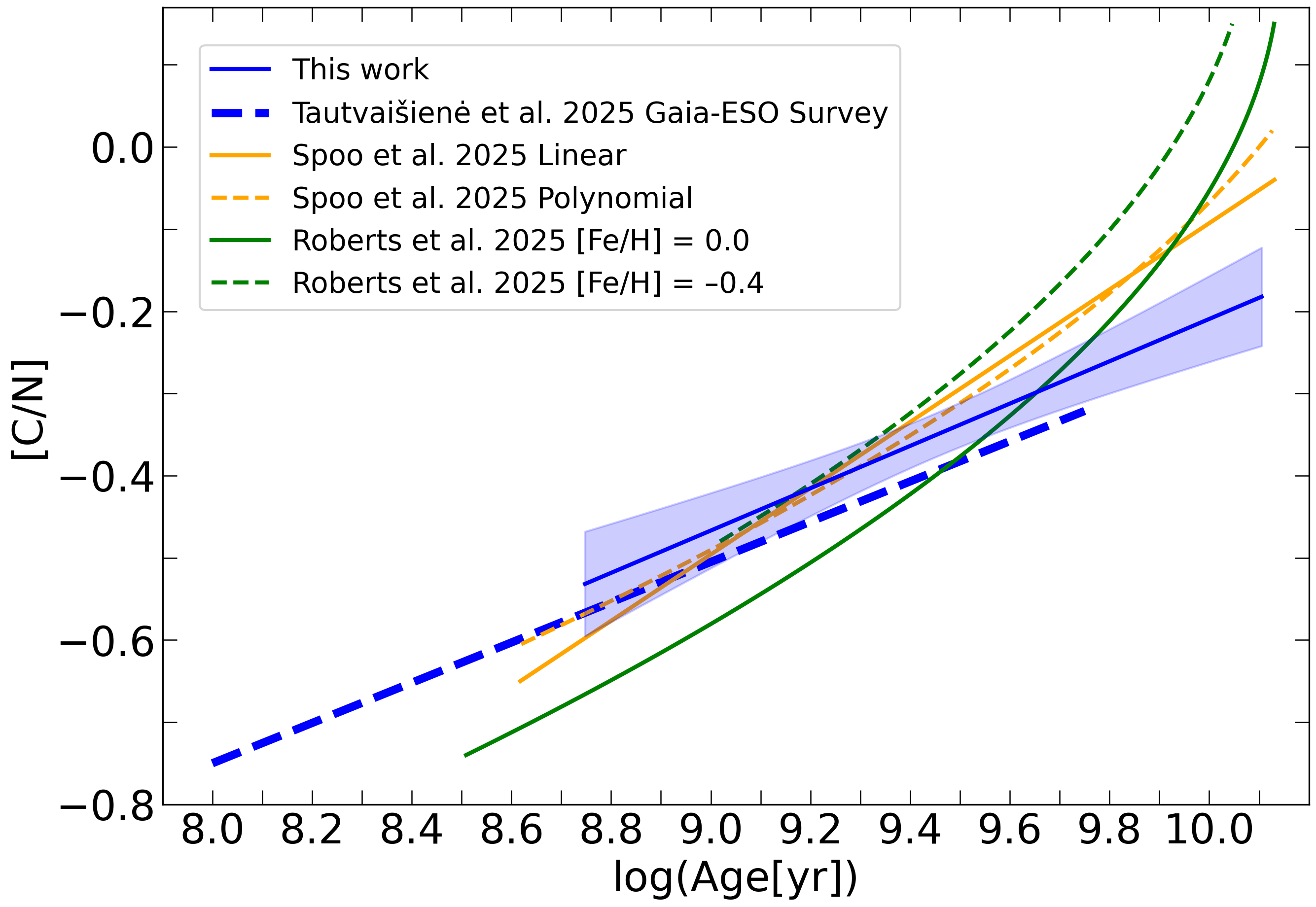}
    \caption{Comparison of the recent [C/N] versus age relations for the 1DUP RGB stars. The blue line and the shadowed confidence interval of 95\% show the relation computed in our work, the thick dashed line is from the $Gaia$-ESO survey open clusters \citep{Tautvaisiene2025}, the orange lines are from \cite{Spoo2025} for the polynomial and linear fits based on open and globular clusters, and the green lines are from \cite{Roberts25} for the field stars of two different metallicities.  }
    \label{CN-age-comparison}
\end{figure}

Recently, \cite{Roberts25} used a much larger sample of stars with [C/N] determined in the Apache Point Observatory Galactic Evolution Experiment (APOGEE) Data Release 17 (\citealt{Abdurro2022}) and used a polynomial function that includes [Fe/H] for the [C/N] relation with age. This approach needs to be further investigated. The fits received by \cite{Roberts25} and our work agree well at log(Age[Gyr])$\sim 9.4$. However, since the [C/N] values in that work even reached $+0.1$~dex for the oldest stars (log(Age[Gyr])$\sim 10$), a polynomial fit was applied. A reason for applying the polynomial fit could be caused by the large number of merged stars in the APOGEE DR17 sample, which according to \cite{Lu2026} have [C/N] values from $-0.05$ -- $0.2$~dex. It would be interesting to check whether such large [C/N] values were also dominated by the thick-disc stars. In our work, the relations were only computed for thin-disc stars. Their [C/N] values at a similar age were not larger than $\sim -0.1$~dex; therefore, the linear relation was sufficient.  In a recent study by \cite{Spoo2025}, both polynomial and linear relations were computed. They included [C/N] values of several metal-deficient globular clusters, and proposed that the same relation could be used in the [Fe/H] interval from $-1.3$~dex to +0.3~dex. However, as we see from the study by \cite{Roberts25}, the [C/N] values in more metal-deficient stars of the same age are higher. The metallicity interval in our study is quite small, but we can infer this tendency as well.  In Fig.~\ref{CN-age-comparison}, we show a comparison of [C/N] versus age relations for 1DUP RGB stars determined in our work and in recent studies by \cite{Roberts25}, \cite{Spoo2025}, and \cite{Tautvaisiene2025}. It is clear that further investigations including the information on the Galactic components, metallicity, and evolutionary stages of stars have to be performed.         

In our work, we also checked the relationship of the [C/N] ratio with the mean galactocentric distance.  As for the [Y/Mg] relation, we assigned stars with $R_{\mathrm{mean}} < 7.5$~kpc to the inner disc, those with $7.5 \le R_{\mathrm{mean}} \le 8.5$~kpc to the solar neighbourhood, and stars with $R_{\mathrm{mean}} > 8.5$~kpc to the outer disc. One may infer that in the inner regions of the Galactic disc, the [C/N] ratios in giant stars are less negative than in the Solar and outer regions because the initial abundances of CNO could be different. This possibility is supported, for example, by investigations of cepheids (\citealt{Luck2011}, \citealt{Luck18}) or H\,{\sc{ii}} regions (\citealt{Arellano2020}). However, as seen from the two bottom panels in Fig.~\ref{CN-age-common}, the uncertainties in the [C/N] determinations and the limited samples of stars do not allow us to see significant differences. In a study of 1475 pre-dredge-up giants from APOGEE DR12 by \cite{Martig16}, no differences in [C/N] values were found in the inner and outer Galactic discs either. This question remains for further studies. We also divided stars of the thin disc according to their maximum distance from the Galactic plane into three groups with |$z_{\rm max}$| from 0 to 0.2~kpc, from 0.2 to 0.4~kpc, and with distances larger than 0.4~kpc, but a negligible difference in the relations of [C/N] with age was found.

\subsection{Comparison of C and N abundances with evolutionary models}
\label{sect:evol}

\begin{figure}
    \centering
    \includegraphics[width=1.0\columnwidth]{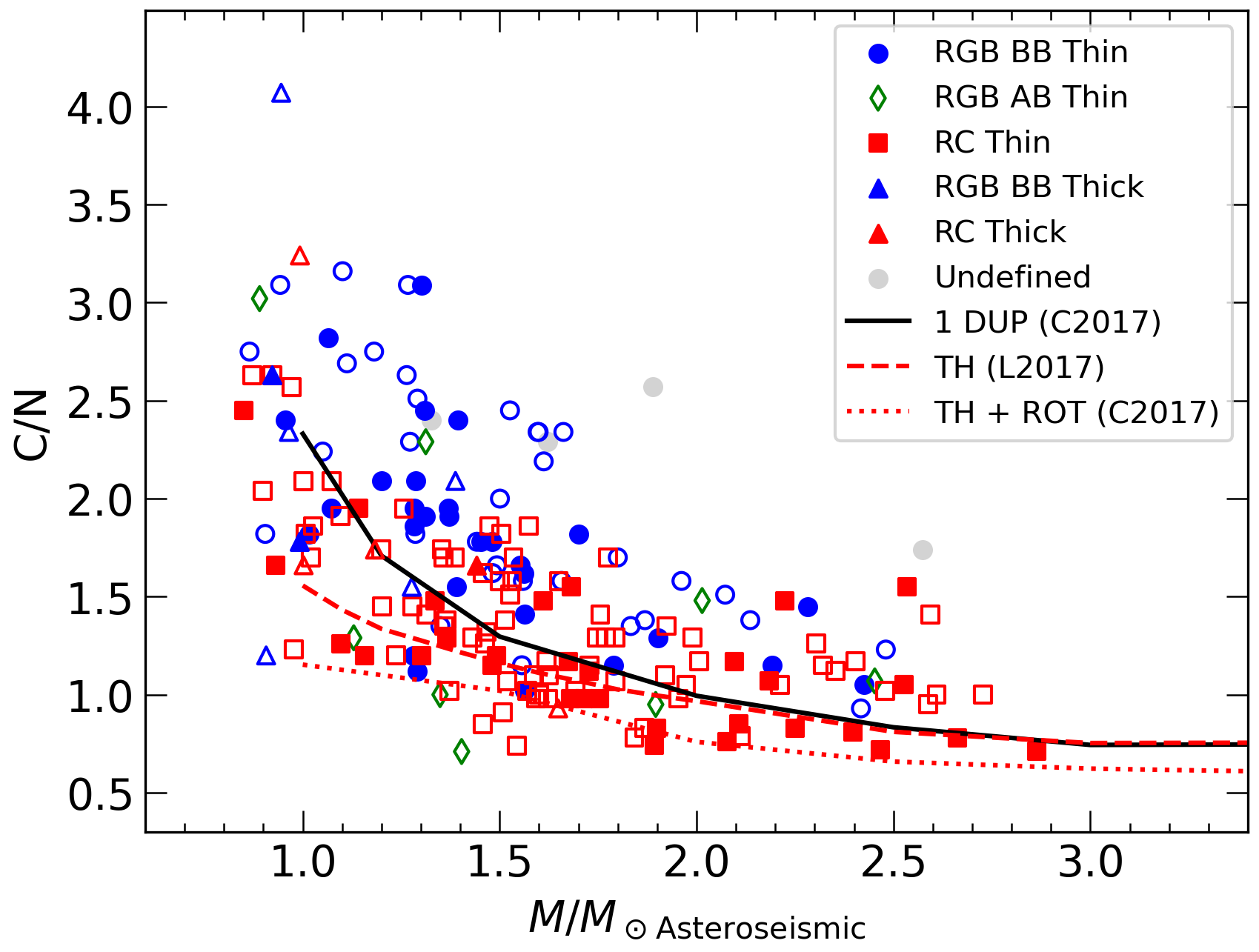}
    \caption{Comparison of C/N ratios of investigated stars with theoretical models for solar-metallicity stars. The filled symbols are for stars with evolutionary stages determined with higher confidence.  The solid black line represents the C/N ratios predicted by the model for stars after the 1\,DUP, taken from \cite{Charbonnel2017}.  The dashed red line represents the minimal values in the model with thermohaline-induced extra mixing (TH), taken from \citet{Lagarde17}. The dotted red line represents the minimal values in the model with thermohaline- and rotation-induced extra mixing (TH+ROT), taken from \citet{Charbonnel2017}.}
    \label{CN-mass}
   \end{figure}

Having a fairly large sample of giants in different evolutionary stages and accurate masses determined using asteroseismic data,  it was interesting to compare the C/N abundance ratios with  theoretical predictions. Fig.~\ref{CN-mass} shows the dependence of the C/N ratio on the stellar mass at different evolutionary stages and models that include alterations of C/N caused by the first dredge-up (\citealt{Charbonnel2017}), as well as by thermohaline-induced extra mixing (\citealt{Lagarde17}) and by thermohaline- and rotation-induced extra mixing (\citealt{Charbonnel2017}). In Fig.~\ref{CN-mass}, we can see that the C/N abundance ratios in the majority of investigated first-ascent RGB stars are affected by the first dredge-up slightly less than predicted by the theoretical model. The rotation-induced extra mixing is also not as efficient as theoretically predicted. The same remarks were made in \cite{Tautvaisiene2025} on the basis of open clusters observed in the $Gaia$-ESO survey. 

\section{Summary and conclusions}

In this work, we searched for solar-type pulsations in a sample of 1250 F, G, and K spectral type giants with $V < 8$~mag in a field of about 45~deg radius centred on the TESS continuous viewing zone in the Northern Hemisphere and derived the asteroseismic ages for 218 of them using PARAM and BASTA codes. This sample was used to investigate [Y/Mg] and [C/N] relations with age, taking into account stellar evolutionary stages and the mean galactocentric distances. In this work, the NLTE Mg and Y abundances were used.      

Our results can be summarised as follows: 

\begin{itemize}

\item 
The [Y/Mg]--age relation exhibits a clear radial dependence across the Galactic disc, with a steeper trend in the outer disc and progressively flatter relations toward the inner disc and the thick disc. In addition, systematic zero-point offsets in [Y/Mg] at fixed age are observed, with the outer disc showing higher [Y/Mg] values than the Solar region, followed by the inner disc and finally the thick disc. These findings highlight differences in star-formation efficiency and chemical enrichment histories. The results are consistent with previous findings of radial variations in $s$-process--to--$\alpha$-element abundance ratios, as reported for [Y/$\alpha$] in open clusters by \citet{Vazquez22} and for [$s$-/Mg] in large samples of disc red giants and main-sequence turn-off and sub-giant-branch stars by \citet{Ratcliffe24}, and confirming the conclusion of \citetalias{Tautvaisiene2021} that [Y/Mg] is not a reliable age indicator for thick-disc stars.

\item 
NLTE abundances of Mg, and especially of Y, have to be used to obtain a more precise stellar age evaluation from [Y/Mg] ratios. 

\item
When using [C/N] abundance ratios as stellar age indicators
for the first-ascent giant stars and helium-core-burning
stars, separate relations have to be used.

\item
The C/N abundance ratios in the majority of investigated first-ascent RGB stars are affected by the first dredge-up slightly less than predicted by the theoretical models. The rotation-induced extra mixing also is not as efficient as theoretically predicted.

\end{itemize}

Considering that reliable solar-type pulsations were identified and asteroseismic ages determined for less than 20\% of giants in the sample of 1250 stars investigated
in this work, methods of alternative stellar age determinations, including chemical clocks, have to be further developed. 

\section*{Data availability}

The full version of Table~\ref{table:Results} is available at the CDS via anonymous ftp to cdsarc.u-strasbg.fr (130.79.128.5) or via http://cdsweb.u-strasbg.fr/cgi-bin/qcat?J/A+A/.

\begin{acknowledgements}

This paper includes data collected by the TESS mission. Funding for the TESS mission is provided by the NASA's Science Mission Directorate.
We acknowledge funding from the Research Council of Lithuania (LMTLT, grant No. S-MIP-23-24). We thank the anonymous referee for helpful suggestions.
\end{acknowledgements}

\bibliographystyle{aa} 
\bibliography{biblio.bib} 

\onecolumn
\begin{appendix}

\section{Machine readable tables of results}

 \begin{longtable}{llll}
 \caption{{\bf Parameters of stars} }
 \label{table:Results}\\
 \hline
 \hline
 Col & Label & Units & Explanations\\
 \hline
 1      & ID                 & --          & Tycho-2 catalogue identification\\
2       & Age\_PARAM          & Gyr         & Asteroseismic age determined using  PARAM(v.1.5) \\
3       & e\_lo\_Age\_PARAM        & Gyr         & Lower uncertainty (68$\%$ credible interval) in the age determination with PARAM \\
4       & e\_up\_Age\_PARAM        & Gyr         & Upper uncertainty (68$\%$ credible interval) in the age determination with PARAM \\
5       & Age\_BASTA          & Gyr         & Asteroseismic age determined using BASTA \\
6       & e\_lo\_Age\_BASTA        & Gyr         & Lower uncertainty (16th quantile) in the age determination with BASTA \\
7       & e\_up\_Age\_BASTA        & Gyr        & Upper uncertainty (84th quantile) in the age determination with BASTA \\
8       & Age\_Asteroseismic  & Gyr         & Asteroseismic age, averaged when both PARAM and BASTA age values available \\
9       &  $\nu_{\rm{max}}$  &     $\mu$Hz           & Frequency at maximum power \\
10       &  $e$\_$\nu_{\rm{max}}$   &    $\mu$Hz      & Uncertainty in frequency at maximum power \\
11       &  $\Delta\nu$       &     $\mu$Hz       & Large frequency separation \\
12      &  $e$\_$\Delta\nu$    &  $\mu$Hz & Uncertainty in large frequency separation \\
13    & Mass\_PARAM              & $M_\odot$    & Stellar mass determined using  PARAM(v.1.5) \\
14    & $e$\_lo\_Mass\_PARAM         & $M_\odot$    & Lower uncertainty (68$\%$ credible interval) of mass determined with PARAM \\
15    & $e$\_up\_Mass\_PARAM         & $M_\odot$    & Upper uncertainty of mass (68$\%$ credible interval) determined with PARAM \\
16    & Mass\_BASTA              & $M_\odot$    & Stellar mass determined with BASTA \\
17    & $e$\_lo\_Mass\_BASTA         & $M_\odot$    & Lower uncertainty (16th quantile) of mass determined with BASTA \\
18    & $e$\_up\_Mass\_BASTA         & $M_\odot$    & Upper uncertainty (84th quantile) of mass determined with BASTA \\
19       & Mass\_Asteroseismic  &    $M_\odot$       & Asteroseismic mass, averaged when both PARAM and BASTA mass values available \\
20     & [Mg/H]LTE                  & dex    & LTE magnesium abundance from \citetalias{Tautvaisiene2020} or \citetalias{Tautvaisiene2022}\\
 21     & [Mg/H]NLTE                  & dex              & NLTE magnesium abundance\\
22     & [Y/H]LTE\_1                  & dex    & LTE yttrium abundance from \citetalias{Tautvaisiene2021}\\
23     & [Y/H]LTE\_2                  & dex    & LTE yttrium abundance determined in this work \\
24     & $e$\_[Y/H]LTE\_2                  & dex    & Uncertainty in LTE yttrium abundance determined in this work \\
25     & [Y/H]NLTE                  & dex              & NLTE yttrium abundance\\
26     & [C/N]                  & dex              & [C/N] ratio from \citetalias{Tautvaisiene2020} or \citetalias{Tautvaisiene2022}\\
27     & C/N               & --           & Carbon-to-nitrogen abundance ratio from \citetalias{Tautvaisiene2020} or \citetalias{Tautvaisiene2022}\\
28    & Evol              & --    & Evolutionary stage (BB or AB -- stars below or above the RGB luminosity bump,\\
        &   &  & respectively. RC -- Clump stars)\\
29    & Evol method       & -- & Method of the evolutionary stage determination. 1 -- if determined purely from \\ 
       &       &  & asteroseismic data, 2 -- if isochronal information was involved \\
30     & Disc  & -- & 0 -- Thin disc, 1 -- Thick disc; taken from \citetalias{Tautvaisiene2020} or \citetalias{Tautvaisiene2022}\\ 
 \hline
 \end{longtable}
 \tablefoot{Full table is available at the CDS.}
 \centering
\end{appendix}

\end{document}